\newif\ifdraft

\documentclass[final,1p,sort&compress]{elsarticle}

\usepackage{graphicx}
\usepackage{bm}
\usepackage{srcltx}
\usepackage{amsmath}
\usepackage{amssymb}
\usepackage{color}
\usepackage{array}
\usepackage{afterpage}

\ifx\pdfoutput\undefined \else \usepackage{epstopdf}\fi

\ifdraft
\usepackage{svn}
\SVN $Id: Lagrangian-particle-algorithms.tex 157 2012-12-03 18:48:00Z shadwick $

\iftrue
\usepackage{showlabels}

\fi

\fi

\newcommand{\be}{\begin{equation}}
\newcommand{\ee}{\end{equation}}
\newcommand{\bea}{\begin{eqnarray}}
\newcommand{\eea}{\end{eqnarray}}

\def\LL{\mathcal{L}}
\def\Lkin{\LL_{\rm kin}}
\def\Lint{\LL_{\rm int}}
\def\Lfield{\LL_{\rm field}}
\def\Lp{\LL_{\mathrm P}}
\def\Lf{\LL_{\mathrm F}}

\def\V{\varphi}
\def\Vt{\widetilde\V}
\def\Ft{\widetilde F}
\def\Gt{\widetilde G}
 
\def\vpot{\Upsilon}

\DeclareMathOperator{\Imag}{Im}

\DeclareMathOperator{\sinc}{sinc}
\DeclareMathOperator{\Dt}{D_t}
\def\Dt{\mathop{D_t}\nolimits}

\def\ma{m_s}
\def\qa{q_s}

\def\me{m_e}
\def\qe{q_e}

\def\qi{q_{\scriptscriptstyle\textsc{I} }}
\def\rhoi{\rho^{\scriptscriptstyle\textsc{(Ion)}}}
\def\nion{n^{\scriptscriptstyle\textsc{(Ion)}}}

\def\Np{N_p}
\def\Ng{N_g}

\def\feL{\Psi^{\scriptscriptstyle(1)}}
\def\K{\mathcal{K}}

\def\bv{\Phi}
\def\bvd{\Phi^\dag}

\def\sump{\sum_{\alpha=1}^{\Np}}
\def\sumg#1{\sum_{#1=1}^{\Ng}}
\def\sumgi{\sumg i}
\def\sumgij{\sumg{i,j}}
\def\sumk{\sum_{k>0}}

\def\summ{\sum_{m=1}^M}
\def\sumbi{\sum_{i=1}^{N_b}}
\def\sumbj{\sum_{j=1}^{N_b}}
\def\sumbij{\sum_{i,j=1}^{N_b}}

\def\xid{\dot\xi}
\def\xidd{\ddot\xi}
\def\xia{\xi_\alpha}
\def\xida{\xid_\alpha}
\def\xidda{\xidd_\alpha}
\def\pid{\dot\pi}
\def\pia{\pi_\alpha}
\def\pida{\pid_\alpha}

\def\xit{\tilde\xi}
\def\xita{\xit_\alpha}

\def\wa{w_\alpha}
\def\fa{f_\alpha}

\def\xt{\tilde x}
\def\vt{\tilde v}

\def\meas#1{d#1}

\def\dmut{\meas\xt\,\meas\vt}
\def\dmu{\meas x\,\meas v}
\def\dxp{\meas x\,\meas p}

\def\intdmut{\intD\dmut}

\def\intdxp{\intD\dxp}

\def\intdx{\intD{\meas x}}
\def\intdp{\intD{\meas p}}
\def\intdv{\intD{\meas v}}

\def\intdxL{\intDl 0L{\meas x}}

\def\intD#1{\int\!\! #1\:}
\def\intDl#1#2#3{\int_{#1}^{#2}\!\! #3\:}

\def\pb#1#2{\left[{#1},{#2}\right]}
\def\PB#1,#2;{\{{#1},{#2}\}}
\def\fd#1,#2;{\frac{\delta{#1}}{\delta{#2}}}

\def\eqref#1{(\ref{#1})}

\allowdisplaybreaks[4]

\biboptions{sort&compress}
\journal{YJCPH}

\begin{document}
\begin{frontmatter}

\ifdraft
\tnotetext[vers]{Version: \SVNId}
\fi

\title{Variational formulation of particle algorithms for kinetic plasma simulations
\ifdraft\tnoteref{vers}\fi}

\author[UNL]{E.~G.\ Evstatiev\corref{cor1}\fnref{fn1}}
\ead{evstati@physics.utexas.edu}
\author[UNL]{B.~A.\ Shadwick}
\ead{shadwick@mailaps.org}

\address[UNL]{Department of Physics and Astronomy, University of Nebraska-Lincoln, NE 68588-0111}

\cortext[cor1]{Corresponding author: E.~G.\ Evstatiev}
\fntext[fn1]{Present address: FAR-TECH, Inc., 10350 Science Center Drive, Bldg. 14, Suite 150, San Diego, CA 92121.}
\date{\today}

\begin{abstract}
Common time-explicit numerical methods for kinetic simulations of plasmas in the low-collisions limit fall into two classes of algorithms: momentum conserving [also known as Particle-In-Cell (PIC)]
and energy conserving.  Each has certain drawbacks.  The PIC algorithm does not conserve total energy, which may lead to spurious numerical heating (grid heating).  Its overall accuracy is at most
second due to the nature of the force interpolation between grid and particle position.  Energy-conserving algorithms do not exhibit grid heating, but because their formulation uses potentials,
computationally undesirable matrix inversions may be necessary.  In addition, compared to PIC algorithms for the same accuracy, these algorithms have higher numerical noise due to the restricted
choice of particle shapes.
Here we formulate time-explicit, finite-size particle algorithms using particular reductions of the particle distribution function.  These reductions are used in two variational principles, a
Lagrangian-based and a Hamiltonian-based in conjunction with a non-canonical Poisson bracket.  The Lagrangian formulations here generalize previous such formulations.  The Hamiltonian formulation is
presented here for the first time.  Many drawbacks of the two classes of particle methods are mitigated.  For example, restrictions on particle shapes are relaxed in energy conserving algorithms,
which allows to decrease the numerical noise in these methods.  The Hamiltonian formulation of particle algorithms is done in terms of fields instead of potentials, thus avoiding solving Poisson's
equation.  An algorithm that conserves both energy and momentum is presented.  Other features of the algorithms include a natural way to perform coordinate transformations, the use of various time
integrating methods, and the ability to increase the overall accuracy beyond second order, including all generalizations.  For simplicity, we restrict our discussion to one-dimensional,
non-relativistic, unmagnetized, electrostatic plasmas.
\end{abstract}

\end{frontmatter}

\section{Introduction}
\label{intro}

Particle-based simulation methods have a long and successful history in plasma physics.  The original proposal by Hockney~\cite{hockney:1826}, based on a method developed for fluid simulations
\cite{harlow:1964}, used computational macro-particles moving in a self-consistent, mean field.  Fields were approximated on a spatial grid and interpolated to the particle position to determine the
Lorentz force.  The essential physics was successfully captured but the simulations suffered from high numerical noise as $\delta$-functions were used to represent the macro-particles.  It was later
realized~\cite{dawson:1983:403}, that $\delta$-function particles used in this way can lead to numerical instability~\cite{Dawson60a}.  A significant improvement was achieved by allowing the
macro-particles to have a finite spatial extent~\cite{langdon_birdsall:2115}; these improved schemes originated the class of methods now known as Particle-In-Cell (PIC) algorithms.  The PIC algorithm
was first used to model plasmas with negligible collisions but later Coulomb collisions and atomic physics were included with techniques based on the Monte Carlo method~\cite{Vahedi:1995}.  PIC
simulations are now widely used to study far ranging plasma systems including modeling laser-plasma interactions~\cite{faure:2004,geddes:2008,yin:2009,Mori:2010aa,vay:2011}, Z-pinches
\cite{welch:2011}, astrophysical and magnetized plasmas~\cite{daughton_kinetic_2005a,lapenta_brackbill:2006:055904,brackbill_lapenta:2008:433}, plasma discharges and low temperature plasma
processing~\cite{Vahedi:1995,Nanbu:2000}, and numerous other applications.

PIC methods have subsequently undergone a significant development.  Research on improvements in reliability and stability lead to the discovery of non-physical, purely numerical artifacts.  The PIC
algorithm does not conserve total energy exactly (even in the absence temporal discretization), which leads to a surprising phenomenon known as ``grid heating" \cite{Langdon:1970aa,okuda:1972:475}.
(See Ref.~\cite{Cormier-Michel:2008bs} for a overview of grid heating.)  Grid heating is attributed to a kinetic instability where sub-grid structures are aliased to low frequency modes (due to finite
grid resolution).  Typically this instability saturates once the plasma has heated to the point where the Debye length is on the order of the grid spacing~\cite{Langdon:1970aa}.  More recently, it has
been shown that the choice of particle shape (the spline used for current and charge deposition and force interpolation) can lead to unphysical effects, especially when considering threshold phenomena
such as self-trapping in a laser-plasma accelerator~\cite{Cormier-Michel:2008bs}.  A further limitation of the PIC algorithm is that its overall accuracy is at most second order in both space and
time.  This is due to the interpolation (typically splines) of quantities between the continuous particle position and the spatial grid.

In an attempt to correct for these deficiencies of the standard PIC algorithm, energy-conserving particle algorithms were devised \cite{lewis:1970:136}.  While strict energy conservation eliminated
the grid heating instability, these algorithms had their own drawbacks.  They did not seem to have the same flexibility with respect to a choice of particle shapes as PIC algorithms, which in turn
affected the level of numerical noise; \textit{i.e.}, for the same numerical accuracy, PIC algorithms had lower noise levels.  Energy conserving algorithms also may require mass-matrix inversions,
which is avoided by the field-based formulations of PIC. As a result, energy-conserving algorithms did not become as popular as PIC algorithms.

One major difference between PIC and energy-conserving algorithms lies in the way each is formulated.  In PIC algorithms \cite{Hockney88,Birdsall:1991aa}, relations between the discretized electric
(vector) potential, electric (magnetic) field, charge deposition, and current deposition were obtained by discretizing the corresponding continuous relations and equations.  In this process, critical
terms of the order of the accuracy of discretization are dropped (as is justified in an asymptotic procedure), but which led to the loss of energy conservation (and possibly violation of other
conservation laws).  In comparison, energy conserving algorithms were derived from a variational principle \cite{lewis:1970:136,Eastwood:1991aa}, using the fact that the Vlasov equation could be
obtained from Low's Lagrangian~\cite{low:1958:282}.  A number of benefits in using variational principle was pointed out: basic properties of the original system were retained in the reduced system; a
natural way of making coordinate transformations was provided; and use of high accuracy space and time solvers was possible.

The goal of this paper is to generalize previous variational formulations and to offer a new formulation of energy conserving algorithms based on the Hamiltonian and the non-canonical Poisson bracket
proposed by Morrison \cite{morrison:1980:383,Weinstein-Morrison81,Morrison:1982aa}.  Our approach is to use particular reductions of the distribution function to a finite collection of terms as well
as particular reductions of the continuous fields to finite number of degrees-of-freedom, either in the Lagrangian or in the Hamiltonian and Poisson bracket.  As a result of our general method, we
show how to avoid many of the previous drawbacks and deficiencies of both PIC and the energy-conserving methods.

In addition to the energy-conserving property of all algorithms in this work, we:
(i)~show that particle shapes in energy conserving algorithms can be chosen with more freedom instead of being a delta-function in space.  In fact, the shape of an extended space particle has very few
physical constraints; the shape may be symmetric about its centroid or may exploit the spatial symmetry of a particular physical problem, \textit{e.g.}, systems with azimuthal symmetry or symmetries in the
gyrokinetic approximation \cite{lee:2003:3196,yulin:2005};
(ii)~relax the method of time integration of the equations of motion (most often leapfrog previously) allowing for higher than second order accuracy.  The choice of a time integrator becomes limited
only by numerical stability.  In this paper we emphasize the spatial discretization and leave time continuous, which allows for convenient formulations of time-explicit schemes.  We prove conservation
laws with continuous time and only in the last step do we choose a particular (explicit) time advancing scheme;
(iii)~show how to reduce continuous quantities with either grid-based reduction (using finite differences) or truncated bases; previous authors have only used truncated bases, which for the case of
finite elements may necessitate mass matrix inversions.  By using grid-based reduction, mass matrices do not appear (for an example, see \ref{HybridModel}), which may be computationally advantageous;
(iv)~derive formulations in terms of fields that are based on the Hamiltonian and a non-canonical Poisson bracket.  Previously only potential-based energy-conserving algorithms have been derived from
a variational principle~\cite{lewis:1970:136,Eastwood:1991aa}.  Using field-based formulations eliminates the need for solving Poisson's equation; and
(v)~derive a particle algorithm that conserves both total energy {\it and} total momentum.  The arguments leading to this algorithm demonstrate the usefulness of the variational approach and exploit
the relations between conservation laws and symmetries of the Lagrangian.  Previous such models were derived by assuming a delta function for the particle shape \cite{evstatiev:2003,evstatiev:2005}.
Throughout, we restrict our discussion to the case of a one-dimensional, nonrelativistic, unmagnetized, electrostatic plasma.  This is done purely to streamline the discussion, elucidating the
central ideas.  The extension of these concepts to the fully relativistic, electromagnetic case, while involving numerous technical details, is largely straightforward and will be the subject of a
future publication.

It has been emphasized before \cite{lewis:1970:136,Eastwood:1991aa} that variational formulations naturally lead to higher overall accuracy particle algorithms (\textit{i.e.}, both in space and
time).  Here we show that this remains true with all of the above relaxed conditions.  We show that force interpolation, field integrators, and time integrators may be chosen to increase the accuracy
of a particle method beyond second order.  In the course of all derivations, we point out where a certain property is being relaxed or is being lost.

Time-implicit formulations of particle algorithms have significant attraction in simulating problems where long time evolution is necessary.  Recently, authors have been successfully reformulated the
PIC algorithm in terms of implicit time integration with the added benefit of energy conservation \cite{chen_chacon:2011,markidis_lapenta:2011}.  However, these formulations do not offer a general
derivation and so is unclear how they can be extended.  For example, these formulations use the Crank-Nicholson time integrator and charge-conserving particle shapes but it is an open question how to
extend these methods to use more accurate time integrators.

While continuous-time formulations are the focus of this paper, we note that one may discretize the action principle in both space and time.  In the context of particle methods for plasma simulations,
this was first considered by Eastwood~\cite{Eastwood:1991aa}.  While Eastwood's method conserved energy exactly, it did so at the expense of an implicit time-advance.  When the time-advance was
altered to be fully explicit, exact energy conservation was lost.  The notion of performing the temporal discretization in the action has been studied extensively by Marsden and co-workers (see for
example Refs~\cite{wendlandt_marsden:1997} and \cite{Marsden:2001aa}).  There are a number of attractive features of this approach and it is a natural extension of the methods presented herein; this
will be the subject of future work.\label{discrete-mechanics}

The paper is organized as follows.  Section~\ref{Lagrangian_PIC} is devoted to deriving algorithms based on a Lagrangian formulation of the Vlasov--Poisson system.  This section presents the finite
differencing formulation.  In section~\ref{TruncatedBasisDerivation} the particle models are derived from a Lagrangian formulation in terms of truncated bases.  The important model which conserves
both momentum and energy is derived there.  In these derivations the fields are described in terms of electrostatic potential.  Section~\ref{BracketDerivation} presents the Hamiltonian derivation of
particle models using truncated basis and a reduction of the non-canonical Poisson bracket.  The equations of motion are formulated in terms of electric field.  Section~\ref{Examples} illustrates
properties of the derived particle models with numerical examples.  Conclusions are in section~\ref{Conclusions}.  \ref{particle-shapes} gives many examples of charge deposition rules, while
\ref{HybridModel} presents a hybrid cold fluid-kinetic particle model from a Lagrangian starting point.  It demonstrates our general method with a different reduction of the particle distribution
function.  It also illustrates how the mass matrix and its inverse may be avoided by the use of grid-based reduction of the continuous quantities.

\section{Lagrangian formulation}
\label{Lagrangian_PIC}

A plasma with negligible collisions is well described by a single-particle phase-space distribution function, $f$, whose phase-space evolution is governed by the Vlasov equation~\cite{Krall:1973aa}
\begin{equation}
	\frac{\partial f}{\partial t} + v \frac{\partial f}{\partial x} + \frac{\qa}{\ma}\,E\,\frac{\partial f}{\partial v} = 0,\label{Vlasov-v}
\end{equation}
where $E=-\nabla\V$ is the electric field, $\V$ is the electric potential, $\ma$ and $\qa$ are the species mass and charge.
  For an initial
phase-space distribution $f_0(\xt, \vt)$, the distribution at any later time is given by
\begin{equation}
	f(x, v, t) = f_0(\xt, \vt),
	\label{f-evolution}
\end{equation}
where $x(t; \xt, \vt)$ and $v(t; \xt, \vt) = \partial x(t; \xt, \vt)/\partial t$ are the macro-particle trajectories with initial conditions $\xt$ and $\vt$: $x(0;\xt,\vt) = \xt$ and $v(0;\xt,\vt) =
\vt$.  The particle trajectories correspond to characteristics of the Vlasov equation and~\eqref{f-evolution} is simply the statement that the distribution function is constant on the characteristics.
Vlasov dynamics can be obtained from the Lagrangian \cite{low:1958:282,Galloway:1971aa,Ye:1992aa}
\begin{multline}
	\LL = \frac{\ma}{2}\intdmut f_0(\xt, \vt)\, \left[\frac{\partial x(t; \xt, \vt)}{\partial t} \right]^2 \\ - \qa\intdmut\, f_0(\xt, \vt)\,\V\left( x(t; \xt, \vt)^{\!\!\!\phantom{k}},t\right)
	+ \frac{1}{8\pi} \intdx \left[ \nabla \V(x) \right]^2,
	\label{Lows_Lagrangian}
\end{multline}
where $x(t; \xt, \vt)$ and $\V(x)$ are to be varied independently. (Here we consider a single-species plasma but the extension to multiple species is
obvious.) 
In the usual way, demanding the action be stationary with respect to variations of the dynamical
variables leads to the equations of motion.  Variation with respect to particle positions gives
\begin{equation}
	\ma\,\ddot x = -\qa\nabla\V\,,
\end{equation}
while variation with respect to the potential gives
\begin{equation}
	\nabla^2\V = -4\pi\,\qa\intdv f(x,v,t) \equiv -4\pi\rho(x).
	\label{Poisson}
\end{equation}
We have used \cite{Galloway:1971aa}
\begin{equation}
	\dmut f_0(\xt,\vt) = \dmu f(x, v, t)
	\label{measure}
\end{equation}
in~\eqref{Poisson} and we have assumed either periodic boundary conditions or an infinite system to allow surface terms to be dropped.  Note that \eqref{measure} is a
statement of particle number conservation and is equivalent to Gardner's restacking theorem \cite{Gardner:1963rz}.

The basic idea of Lagrangian macro-particle methods lies in the representation of the full distribution function $f({ x}, { v}, t)$ as a sum of moving spatial volumes, $\fa({ x}, {v}, t)$, called
macro-particles:
\begin{eqnarray}
	f({ x}, {v}, t) &=& \sum_\alpha \fa({ x}, {v}, t) \nonumber \\
	&=& \sum_\alpha\wa\,S[{ x}-\xia(t)] \,\delta[{ v}-\xida(t)].
	\label{Sum_f_i}
\end{eqnarray}
The choice of a delta function in velocity space is not essential but avoids the necessity to track stretching phase space volumes, which is why we adhere to it.  In Eq.~(\ref{Sum_f_i}), $\wa$ are
constant weights and the function $S$ is the fixed spatial extent of the computational particle (hereafter we use the terms particle, computational particle, and macro-particle interchangeably unless
otherwise specified) and is normalized as
\begin{equation}\label{Shape_norm}
	\intdx S[{ x}-\xia(t)] = 1.
\end{equation}
An additional simplification is made by assuming that all particles have the same shape.  We note that the representation (\ref{Sum_f_i}) is general and independent of whether both electric and
magnetic fields are present in the system, {\it i.e.}, it is valid for the general Vlasov--Maxwell system.  For clarity of the presentation, in this paper we consider only electrostatic,
non-relativistic models.  Electromagnetic and relativistic models can be derived similarly to this presentation and will be presented in a future publication.  We view \eqref{Sum_f_i} as a particular
\emph{reduction} of the particle distribution function.  \ref{HybridModel} gives another example of such a reduction.

Substituting our form of the distribution function, \eqref{Sum_f_i}, into the Lagrangian and again using \eqref{measure}, we obtain a reduced Lagrangian
\begin{equation}
	\begin{aligned}
	\LL &= \frac\ma2\sump \wa\,\xid^2_\alpha - \qa \sump \wa \intdx S(x-\xia)\,\V(x) + \frac{1}{8\pi}\intdx\left(\nabla\V\right)^2\\
	&= \Lkin + \Lint + \Lfield\,,
	\end{aligned}
	\label{L_Cont}
\end{equation}
where
\begin{align}
\Lkin &= \frac\ma2\sump \wa\,\xida^2, \label{L_kin} \\ 
\Lint &= \intdx S(x-\xia)\,\V(x)\,, \label{coupling} \\ 
\Lfield &= \frac1{8\pi}\intdx\left(\nabla\V\right)^2.\label{field}
\end{align}
Although we have replaced a continuum of particles with labels $\xt$ and $\vt$ by $\Np$ macro-particles, we still have an infinite degree-of-freedom system due to the presence of the continuous field
$\V$.
The equations of motion are obtained from \eqref{L_Cont} by considering variations of the particle position and of the potential.  For the particles, the usual Euler--Lagrange equation
\begin{equation}
	\frac d{dt}\,\frac{\partial\LL}{\partial\xida} - \frac{\partial\LL}{\partial\xia} =0,
	\label{EOM_xi_alpha_temp}
\end{equation}
gives
\begin{equation}
	\xidda = - \frac\qa\ma\intdx\frac{\partial S}{\partial\xia} \V(x) = -\frac\qa\ma\intdx S[{x}-\xia(t)] \nabla \V\,.
	\label{xi-EOM-Cont}
\end{equation}
Since the potential is a field, the Euler--Lagrange equation for the potential is
\begin{equation}
	\frac{\delta\LL}{\delta \V} = 0, \, 
\end{equation}
where $\delta/\delta \V$ denotes a functional derivative.  Then
\begin{equation}
	\nabla^2 \V = - 4\pi\,\qa\sump \wa S[{ x}-\xia(t)]\, .
	\label{Poisson-Cont}
\end{equation}
Note that the factor $\qa/\ma$ appearing in \eqref{xi-EOM-Cont} is the physical charge to mass ratio of the plasma species.  It is not necessary to make the ad-hoc assumption that the macro-particle
have the same charge to mass ratio as the plasma species, this is a consequence of the phase-space decomposition \eqref{Sum_f_i}.  Furthermore, the second form of the force in \eqref{xi-EOM-Cont} may
clearly be interpreted as the electric field averaged over the particle shape.

The substitution of \eqref{Sum_f_i} into \eqref{Lows_Lagrangian} is equivalent to a choice of a trial function for $f$ that depends on a number of parameters, which in our case are the particle
positions and velocities.  The values of these parameters are obtained by solving the equations resulting from the variation \eqref{EOM_xi_alpha_temp}.  Other choices of trial functions may lead to
models without particles at all \cite{lewis_barnes:1987,Shadwick:2010aa}.

A significant advantage of the variational formulation is the connection between symmetries and conservation laws as embodied in Noether's theorem \cite{Jose:1998aa}.  Our introduction of macro-particles
through \eqref{Sum_f_i} neither results in explicit time-dependence in the Lagrangian nor breaks translational invariance of the Lagrangian, thus we should expect the equations of motion 
\eqref{xi-EOM-Cont} and \eqref{Poisson-Cont} to exactly conserve both energy and momentum.  The total energy of the system is the sum of macro-particle kinetic energy and field energy, 
\begin{equation}
	W = \frac\ma2\sump \wa\,\xida^2 + \frac{1}{8\pi}\intdx\left(\nabla\V\right)^2.
	\label{W-cont}
\end{equation}
Using the equations of motion, it is straightforward to see that $W$ is an invariant:
\begin{align}
	\frac{dW}{dt} & = \ma\sump \wa\,\xida\,\xidda - \frac{1}{4\pi}\intdx\V\,\frac{\partial}{\partial t}\,\nabla^2\V\nonumber\\
	 & = -\qa\sump \wa\,\xida\intdx S(x - \xia)\nabla\V + \qa\intdx\V\,\frac{\partial}{\partial t}\sump \wa S(x-\xia)\nonumber\\
	 & = -\qa\sump \wa\,\xida\intdx S(x - \xia)\nabla\V - \qa\sump \wa\,\xida\intdx\V\,\frac{\partial}{\partial x}S(x-\xia)\nonumber\\
	 & = -\qa\sump \wa\,\xida\intdx S(x - \xia)\nabla\V + \qa\sump \wa\,\xida\intdx S(x-\xia)\,\nabla\V\nonumber\\
	 & = 0\,.
\end{align}
The total momentum of the system is simply
\begin{equation}
	P = \ma\sump \wa\,\xida\,,
	\label{P_cont}
\end{equation}
since the electrostatic field carries no momentum (the Poynting vector is zero in the electrostatic approximation).  Now
\begin{align}
	\frac{dP}{dt} & = -\qa\sump \wa\intdx S(x-\xia)\nabla\V \nonumber\\
	 & = \frac1{4\,\pi}\intdx \nabla^2\V\nabla\V \nonumber\\
	 & = \frac1{8\,\pi}\intdx \nabla\left(\nabla\V\right)^2 \nonumber\\
	 & = 0\,,
\end{align}
where we have used \eqref{Poisson-Cont}.

At this point in our reduction, we have a finite number of macro particles representing the plasma but a continuous field for the potential.  Here one must provide some approximate or exact solution
of \eqref{Poisson-Cont}, which is then used to integrate the macro-particle equations of motion.  One possibility would be to use methods based on evaluating the Green's function, constructing $\V$ as
the superposition of the potentials due to each macro-particle.  Even though the macro-particles interact via the mean field, computation of $\V$ using Green's functions scales as $O(\Np^2)$ and is
thus limited to relatively small systems.  A more computationally advantageous alternative is to introduce a discrete representation for the potential.  There are two general approaches: using a
spatial grid, approximating the potential by its value at the grid point, or using a truncated set of (local or global) basis functions and representing the potential by its projection onto the basis.

The interaction term in the Lagrangian, \eqref{coupling}, provides both the force in \eqref{xi-EOM-Cont} as well as the charge density in \eqref{Poisson-Cont}.  Of course, this will continue to be the
case when the continuous potential is replaced by a discrete approximation.  It will be necessary to approximate $\Lint$ consistently with the choice of the discrete potential but this single
approximation ultimately yields both the force term in the $\xia$ equation of motion and the charge density in a discrete analogue of Poisson's equation.  Thus we are guaranteed that these terms are
consistently approximated.

\subsection{Discretization using a spatial grid}
\label{spatial-grid}

We assume a fixed spatial grid $x_i$ with $i\in[1,\Ng]$ and grid spacing $h$ with $\V_i$ being the numerical approximation to $\V(x_i)$.  We must now approximate two terms in \eqref{L_Cont}, $\Lint$
and $\Lfield$.  The interaction term requires knowledge of $\V$ between the grid-points; so some manner of interpolation is required.  Finite elements \cite{Becker:1981aa} offer a consistent way to
perform such interpolations to any accuracy.  Let $\Psi_i(x)$, $i = 1,\ldots,\Ng$ be finite-element basis of some order.  We interpolate $\V$ between the grid points by
\begin{equation}
	\V(x) = \sumgi \V_i\Psi_i(x)
	\label{fe-interp}
\end{equation}
and thus \eqref{coupling} becomes
\begin{equation}
	\intdx S(x-\xia)\,\V(x) = \sumgi\V_i\intdx S(x - \xia)\Psi_i(x) = \sumgi\V_i\rho_i(\xia)\,,
	\label{discrete-coupling}
\end{equation}
where
\begin{equation}
	\rho_i(\xia) = \intdx S(x - \xia)\Psi_i(x)
	\label{deposition}
\end{equation}
is the effective (projected) shape of the particle.  Note the expression for $\rho_i$ can be computed analytically since the function $S$ is known.  If $\Psi_i(x)$ are constructed from Lagrange
polynomials, then $\sumgi \Psi_i(x)$ = 1 and
\begin{equation}
	\sumgi \rho_i(\xia) = \sumgi \intdx S(x - \xia)\Psi_i(x) = \intdx S(x - \xia) = 1\,.
	\label{charge-norm}
\end{equation}
This property means that the total charge deposited on the grid and at any instant of time is constant.

It remains to approximate $\Lfield$ in terms of $\V_i$.  This can be approached in two ways which give roughly equivalent results.  We can use \eqref{fe-interp} to write the integral in \eqref{field} as
\begin{equation}
	\intdx\left(\nabla\V\right)^2 = \sumgij\V_i\V_j\intdx\frac{d\Psi_i(x)}{dx}\,\frac{d\Psi_j(x)}{dx}\,.
\end{equation}
Defining
\begin{equation}
	-hK_{ij} = \intdx\frac{d\Psi_i(x)}{dx}\,\frac{d\Psi_j(x)}{dx}\,,
	\label{Kij}
\end{equation}
we have
\begin{equation}
	\intdx\left(\nabla\V\right)^2 = -h\sumgij\V_i\V_jK_{ij}\,.
	\label{field-fe}
\end{equation}
Alternatively, after integrating by parts in $\Lfield$, we can approximate the integral as
\begin{equation}
	\intdx\left(\nabla\V\right)^2 = - \intdx\V(x)\nabla^2\V(x) \approx -h\sumgi \left.\V_i \frac{d^2\V}{dx^2}\right|_{x_i}.
\end{equation}
While this appears to simply be using the trapezoidal rule to evaluate the integrand, with either periodic boundary conditions or an infinite domain, this approximation has spectral accuracy, that
is, all modes supported by the grid are integrated exactly.  We complete the approximation by choosing a finite-difference representation for the second derivative.  Regardless of details of the
finite-difference approximation, it can always be expressed as
\begin{equation}
	\left.\frac{d^2\V}{dx^2}\right|_{x_i} = \sumg j\widetilde K_{ij}\V_j + O(h^a),
	\label{K_tilde}
\end{equation}
for some integer $a$.  Thus we have
\begin{equation}
	\intdx\left(\nabla\V\right)^2 \approx -h\sumgij \V_i \widetilde K_{ij}\V_j\,,
	\label{field-fd}
\end{equation}
which has the same form as \eqref{field-fe}. Notice that while $K_{ij}$ is always symmetric [cf.~\eqref{Kij}], this need not be true for $\widetilde{K}_{ij}$.

We now arrive at the finite degree-of-freedom Lagrangian
\begin{equation}
	\LL = \frac\ma2\sump \wa\,\xid^2_\alpha - \qa \sump\sumgi \wa \rho_i(\xia)\,\V_i - \frac{h}{8\pi}\sumgij\V_i\K_{ij}\V_j ,
	\label{L-grid}
\end{equation}
where we take either $\K_{ij} = K_{ij}$ or $\K_{ij} = \widetilde K_{ij}$.  The dynamical equations are obtained by demanding that the action be stationary with respect to variations in both
$\xia$ and $\V_i$.  Taking these variations yields
\begin{equation}
	\xidda = - \frac\qa\ma\sumgi\frac{\partial\rho_i(\xia)}{\partial\xia}\,\V_i\,
	\label{xi-EOM}
\end{equation}
and
\begin{equation}
	\sumg j\K_{ij} \V_j = - 4\pi\,\frac\qa h\sump \wa\,\rho_i(\xia)\,.
	\label{Poisson-d}
\end{equation}
The reason we choose to integrate by parts in \eqref{field-fd} is now clear: by dong so, we are able to directly specify the difference method for the second derivative that appears in Poisson's
equation, \eqref{Poisson-d}.

Discretizing $\V(x)$ in \eqref{W-cont} in the same manner as in the Lagrangian, we have
\begin{equation}
	W_L = \frac\ma2\sump \wa\,\xida^2 - \frac{h}{8\pi}\sumgij\V_i\K_{ij}\V_j.
	\label{W}
\end{equation}
Using the equations of motion, we find
\begin{align}
	\frac{dW_L}{dt} & = \ma\sump \wa\,\xida\,\xidda - \frac{h}{4\pi}\sumgij\V_i\,\K_{ij}\frac{d\V_j}{dt}\nonumber\\[2pt] & = -\qa\sump\sumgi
	\wa\,\xida\,\V_i\,\frac{\partial\rho_i}{\partial\xia} + \qa\sumgi\V_i\sump \wa\frac{d\rho_i(\xia)}{dt}\nonumber\\[2pt] & = -\qa\sump\sumgi
	\wa\,\V_i\,\frac{\partial\rho_i}{\partial\xia}\,\xida + \qa\sumgi\sump \wa\,\V_i\,\frac{\partial\rho_i}{\partial\xia}\xida\nonumber\\
	 & = 0\,.
\end{align}
Introducing a spatial grid does not affect energy conservation.  This is expected since the spatial discretization of $\V$ does not introduce explicit time-dependence into the Lagrangian.  An
immediate advantage of the variational approach is that models derived in this way are automatically free of grid heating.  A side effect of introducing a spatial
grid is that it breaks the translation invariance of $\LL$ and consequently total momentum is no longer exactly conserved; see section~\ref{TruncatedBasisDerivation} for a more complete discussion.

\begin{figure}[htb]
	\centering
	\includegraphics{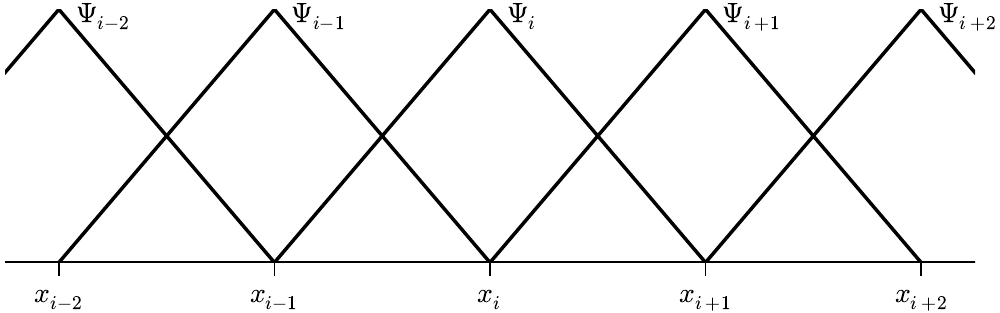}
	\caption{Linear finite element basis functions.  The basis functions $\Psi_i$ are identified with the grid-point at which it takes on the value $1$; e.g., $\Psi_i$ is the tent function with
	support $[x_{i-1},x_{i+1}]$.}
	\label{linearFE}
\end{figure}

We conclude this section by providing a concrete example of this procedure to derive a model that is second-order accurate in $h$.  For simplicity, we consider the case of a charge-neutral electron
plasma with an immobile ionic background and a spatially periodic domain.  Since this system has no ion dynamics, we can forgo summing over species and simply introduce the ion density into the
Lagrangian
\begin{equation}
	\LL = \frac\me2\sump \wa\,\xid^2_\alpha - \qe \sump\sumgi \wa \rho_i(\xia)\,\V_i - \sumgi \rhoi_i\,\V_i - \frac{h}{8\pi}\sumgij\V_i\K_{ij}\V_j,
	\label{L-qn}
\end{equation}
where
\begin{equation}
	\rhoi_i = \qi\intdx \nion(x)\Psi_i(x),
	\end{equation}
with $\nion(x)$ being the given ion density.  Linear finite elements yield second order accuracy interpolation \cite{Becker:1981aa} (see Figure
\ref{linearFE}):
\begin{equation}
	\feL_i(x) = 
	\begin{cases}
		1 - \dfrac{|x - x_i|}h & x_{i-1} \le x \le x_{i+1},\\
		0 &\textrm{otherwise.}
	\end{cases}
\end{equation}
To determine $\rho_i$ we need to specify $S(x)$.  Regardless of the choice of $S$, the accuracy of the interpolation will be second order due to our basis choice.  The choice of $S$ affects the
quality of our approximation through the extent to which \eqref{Sum_f_i} is a good ansatz but has no influence on the formal order of the model.  Arguably the simplest choice for $S$ is a top-hat
cell-wide function:
\begin{equation}
	S(x - \xia) = \begin{cases}\dfrac1h& |x - \xia| \le \frac12\,h,\\[6pt]0 &\textrm{otherwise.}\end{cases}
	\label{top-hat-S}
\end{equation}
While we have chosen $S$ to be exactly one grid-cell wide, this is by no means essential.  The choice of support of $S$ is completely independent of the grid spacing; the particular choice in
\eqref{top-hat-S}, allows us to make connection with the usual PIC particle shapes (see Figure \ref{shapes-fig} and Table \ref{shapes}).

We now use \eqref{deposition} to determine the grid charge deposition.  With this $S$, for any $\xia$, there are only three values of $i$ for which $\rho_i(\xia)\ne 0$.  Take $x_k$ to be the grid
point nearest $\xia$ and let $\Delta = (\xia - x_k)/h$.  Clearly $|\Delta| \le 1/2$.  It is straightforward to evaluate \eqref{deposition} to obtain
\begin{equation}
	\begin{aligned}
		\rho_{k-1}           &= \frac12\left(\Delta - \frac12\right)^2,\\
		\rho_{k\phantom{-1}} &= \frac34-\Delta^2,\\
		\rho_{k+1}           &= \frac12\left(\Delta + \frac12\right)^2.
	\end{aligned}
	\label{quadratic}
\end{equation}
This is equivalent to the charge deposition obtained from the usual PIC quadratic particle shape \cite{Hockney88}.  It is possible to recover all of the usual smooth particle shapes.  For example,
taking 
\begin{equation}
	S(x - \xia) = \frac1h
	\begin{cases}
		\displaystyle \frac{3}{4} - \left(x - \xia\right)^2                   &           |x - \xia| \le \frac12h,\\[8pt]
		\displaystyle \frac12\left(\frac32 - |x - \xia|\right)^2 & \frac12h < |x - \xia| \le \frac32h,\\[8pt]
		\displaystyle 0                                        &\text{otherwise,}
	\end{cases}
	\label{S2}
\end{equation}
we obtain
\begin{equation}
	\begin{aligned}
		\rho_{k-2}           &= \frac1{24}\left(\Delta - \frac12\right)^4\\[8pt]
		\rho_{k-1}           &= \frac{19}{96} - \frac{11}{24}\,\Delta + \frac14\,\Delta^2 + \frac{1}{6}\,\Delta^3 - \frac{1}{6}\,\Delta^4\\[8pt]
		\rho_{k\phantom{-1}} &= \frac{115}{192} - \frac58\,\Delta^2 + \frac14\,\Delta^4\\[8pt]
		\rho_{k+1}           &= \frac{19}{96} + \frac{11}{24}\,\Delta + \frac14\,\Delta^2 - \frac{1}{6}\,\Delta^3 - \frac{1}{6}\Delta^4\\[8pt]
		\rho_{k+2}           &= \frac1{24}\left(\Delta + \frac12\right)^4
	\end{aligned}
	\label{quartic}
\end{equation}
which is equivalent to the usual quartic charge deposition rule.  We take up the matter of particle shapes in some detail in \ref{particle-shapes}.  In particular, we demonstrate a cubic $\rho_{k}$
spanning only three grid-points.

All that remains is to approximate \eqref{field} to second-order accuracy.  Evaluating \eqref{Kij} for our linear basis, $\feL_i$ is straightforward.  For any $i$, we can see that $\feL_i{}'(x)$ has a
non-zero overlap with $\feL_j{}'(x)$ only for $j= i\pm1$ and thus we have
\begin{equation}
	K_{ij} =
	\begin{cases}
		-\dfrac2{h^2} & j = i,\\[8pt]
		\phantom{-}\dfrac1{h^2} & j = i\pm1\,.
	\end{cases}
\end{equation}

Alternatively, we can use finite difference approximations and evaluate \eqref{field} using \eqref{field-fd}.  Since we are considering a periodic domain, it is reasonable to choose a central 
difference approximation
\begin{equation}
	\frac{d^2\V}{dx^2}\Biggl|_{x_i} = \frac{\V_{i+1} - 2\,\V_i + \V_{i-1}}{h^2} + O(h^2)\,,
\end{equation}
which gives
\begin{equation}
	\widetilde K_{ij} = K_{ij}\,.
\end{equation}
Linear finite-elements give the same approximation to \eqref{field} as taking second-order central differences.  This is essentially a coincidence; higher-order finite element bases (quadratic, cubic,
\textit{etc.}) do not yield expressions for $K_{ij}$ that can be equated to conventional differencing schemes.  Nothing, other than the symmetry of the problem, forces us to choose central differences;
any second-order approximation would suffice.  Note, using an expression for $\widetilde K_{ij}$ that is accurate beyond second-order will not increase the overall spatial order of the method unless a
corresponding more accurate interpolation scheme is used to evaluate \eqref{deposition}.

The macro-particle equation of motion, which is not alerted by the specifics of the spatial discretization, remains as in \eqref{xi-EOM}.  For a uniform ion background, $\nion_0$, we have
\begin{equation}
	\rhoi_i = \qi\,\nion_0\intdx \feL_i(x) = \qi\,\nion_0\,h\,.
\end{equation}
For completeness we restate Poisson's equation including the ionic background and our approximation for $\K_{ij}$
\begin{equation}
	\frac1{h^2}\left(\V_{i+1} - 2\V_i + \V_{i+1}\right) = - 4\pi\,\frac\qa h\sump \wa\,\rho_i(\xia) - 4\pi\,\qi\,\nion_0\,.
\end{equation}

\subsection{Discretization using a truncated basis and the question of momentum conservation}
\label{TruncatedBasisDerivation}

In the previous section the interaction and field parts of the Lagrangian (\ref{L_Cont}) were reduced from an infinite to a finite degree of freedom quantities by discretizing on a grid.  We noted
that the introduction of a spatial grid breaks the translational invariance of $\LL$, which leads to loss of momentum conservation in the reduced system.  In this section we consider a reduction using
a truncated global basis and investigate the question of momentum conservation in this case.  We show that replacing the continuous potential by a finite collection of projections onto a truncated
basis can result in a discrete system that retains translation invariance.  (Of course, if the basis is not truncated, which is not useful from a computational perspective, then we would expect
translation invariance to be maintained for any complete basis.)

Let $\bv_m(x)$, $m = 1,\ldots,
M$ be the first $M$ elements of an orthonormal basis.  We approximate the potential as
\begin{equation}
	\V \approx \summ \V_m\,\bv_m(x),
\end{equation}
where
\begin{equation}
	\V_m = \intdx \bvd_m(x)\,\V(x)
\end{equation}
and $\bvd_m(x)$ is the dual to $\bv_m(x)$ satisfying
\begin{equation}
	\intdx \bv_m(x)\,\bvd_n(x) = \delta_{mn},
	\label{bv-orthog}
\end{equation}
for $m,n = 1,\ldots,M$.  The accuracy of this approximation will depend on the number of elements kept and the convergence properties of the basis.  For a truncated system to possess translation
invariance requires that the basis have certain properties.  These properties are made evident by shifting the origin by an amount $\delta x$ while leaving the physical system unchanged and requiring
this transformation to be a symmetry of $\LL$ \cite{Jose:1998aa}.  The kinetic term, $\Lkin$, is obviously translation invariant under such a shift since $\delta x$ is time independent.  Consider
$\Lint$.  Let $x'$ be the new coordinate, with $x = x' + \delta x$.  The particle coordinates and potential relative to $x'$ are
\begin{equation}
	\begin{aligned}
	  \xita & = \xia - \delta x , \\
		\Vt(x') & = \V(x' + \delta x) \,.
	\end{aligned}
\end{equation}
The symmetry condition is then
\begin{equation}
	\Lint[\V, \xia] = \Lint[\Vt,\xita]\,.
\end{equation}
Introducing our basis expansion into $\Lint$ and suppressing prime symbols, we have
\begin{equation}
	\begin{aligned}
		\Lint[\Vt,\xita] &= -\qa\sump\summ \wa\,\Vt_m\intdx S(x - \xita)\bv_m(x)\\
		&= -\qa\sump\summ \wa\,\Vt_m\intdx S(x - \xia + \delta x)\bv_m(x)\\
		&= -\qa\sump\summ \wa\,\Vt_m\intdx S(x - \xia)\bv_m(x - \delta x)
	\end{aligned}
\end{equation}
and
\begin{equation}
	\Vt_m = \intdx \bvd_m(x)\,\Vt(x) = \intdx \bvd_m(x)\,\V(x+\delta x) = \intdx \bvd_m(x - \delta x)\,\V(x)\,.
\end{equation}
Combining these expressions and expanding to lowest order in $\delta x$, we have
\begin{align}
	\Lint[\Vt,\xita] &=  -\qa\sump\summ \wa\Biggl\{\V_m\intdx S(x - \xia)\bv_m(x)\nonumber\\[6pt]
	&\hskip-36pt - \delta x\Biggl[\intdx\V(x)\,\frac{d\bvd_m(x)}{dx}\intdx S(x - \xia)\bv_m(x)\nonumber\\
	&\hskip115pt+\intdx\V(x)\bvd_m(x)\intdx S(x - \xia)\,\frac{d\bv_m(x)}{dx}\Biggr]\Biggr\}\nonumber\\
	&= \Lint[\V, \xia] +\delta x\,\qa\sump\summ \wa\Biggl[\intdx\V(x)\,\frac{d\bvd_m(x)}{dx}\intdx S(x - \xia)\bv_m(x)\nonumber\\[4pt]
	&\hskip115pt+ \intdx\V(x)\bvd_m(x)\intdx S(x - \xia)\,\frac{d\bv_m(x)}{dx}\Biggr].
	\label{Lint-symm}
\end{align}
Our symmetry condition requires that the term multiplying $\delta x$ in \eqref{Lint-symm} vanish.  For the symmetry to exist independent of the particle shape $S$ and the details of the potential, this
term must vanish for each $m$.  We are led to the condition
\begin{equation}
	\frac{d\bv_m(x)}{dx} = \pm \alpha(m)\,\bv_m(x)\quad\textrm{and}\quad \frac{d\bvd_m(x)}{dx} = \mp\alpha(m)\bvd_m(x).
	\label{m-c-cond}
\end{equation}
On a finite domain with periodic boundary conditions this condition is satisfied by the discrete Fourier basis; we are not aware of any other discrete basis that fulfills \eqref{m-c-cond} on either a
finite or infinite domain. Using the discrete Fourier basis, it is straightforward to show that $\Lfield$ is also translation invariant.

We now specialize our discussion to the case of a truncated Fourier basis.  Let
\begin{equation}
	\begin{aligned}
		\bv_k(x) &= e^{ikx}\\
		\bvd_k(x) & = \frac1L\,e^{-ikx}
	\end{aligned}
	\label{fb-def}
\end{equation}
where $k = 2\,m\,\pi/L$, $m=0,\pm1,\ldots,\pm M$ and $L$ is the domain size.  With this basis, the interaction term becomes
\begin{equation}
	\begin{aligned}
		\Lint & = -\qa\sump\sum_k \wa\V_k\intdxL S(x - \xia)\bv_k(x)\\
		 & = -\qa L\sump\sum_k \wa\V_k\left[\intdxL S(x - \xia)\bvd_k(x)\right]^* \\
		 & =  -\qa L\sump\sum_k \wa\V_k\,\rho^*_k(\xia)\,,
	\end{aligned}
\end{equation}
where
\begin{equation}
	\rho_k(\xia) = \intdxL S(x-\xia)\,\bvd_k(x),\label{rho_k-fb}
\end{equation}
and
\begin{equation}
	\V_k = \intdxL \V(x)\,\bvd_k(x),\label{phi_k-fb}
\end{equation}
and we have used the relation $\bv_k(x)/L = [\bvd_k(x)]^*$.  We also need to evaluate \eqref{field}:
\begin{align}
	\Lfield &= \frac1{8\pi}\sum_{k,k'}\V_k\,\V_{k'}\intdxL\frac{d\bv_k(x)}{dx}\,\frac{d\bv_{k'}(x)}{dx}\nonumber\\
	&= -\frac1{8\pi}\sum_{k,k'}k\,k'\V_k\,\V_{k'}\intdxL\bv_k(x)\bv_{k'}(x)\nonumber\\
	&= -\frac L{8\pi}\sum_{k,k'}k\,k'\V_k\,\V_{k'}\intdxL\bv_k(x)\bigl[\bvd_{k'}(x)\bigr]^*\\
	&= -\frac L{8\pi}\sum_{k,k'}k\,k'\V_k\,\V_{k'}\intdxL\bv_k(x)\bvd_{-k'}(x)\nonumber\\
	&= \frac L{8\pi}\sum_{k}k^2\V_k\,\V_{-k} = \frac L{4\pi}\sumk k^2\V_k\,\V^*_k,\nonumber
\end{align}
where, since $\V$ is real, $\V_{-k} = \V^*_k$.  Finally we arrive at the discrete form of the Lagrangian
\begin{equation}
	\LL = \frac\ma2\sump \wa\,\xid^2_\alpha - \qa\,L\sump\sum_k \wa\,\V_k\,\rho^*_k(\xia) + \frac{L}{4\pi}\sumk k^2\V_k\,\V^*_k\,.
	\label{L-tfb}
\end{equation}

To obtain the equations of motion, we require the action to be stationary with respect to variations of $\xia$ and $\V_k$ (since $\V_k$ and $\V^*_k$ are not independent, we need only consider
variations of $\V_k$).  The equation of motion are
\begin{equation}
	\xidda = - \frac{\qa L}\ma\sum_k\frac{\partial\rho^*_k(\xia)}{\partial\xia}\,\V_k\,,
\end{equation}
and
\begin{equation}
	k^2\V_k = 4\pi\,\qa \sump \wa\,\rho_k(\xia)\,.
	\label{poisson-fb}
\end{equation}
Using \eqref{rho_k-fb} and \eqref{fb-def} it is easy to show
\begin{equation}
	\frac{\partial\rho_k(\xia)}{\partial\xia} = -ik\,\rho_k(\xia)
	\label{drhodxi}
\end{equation}
allowing us to write the equation of motion as
\begin{align}
	\xidda &= - i\,\frac{\qa L}\ma\sum_k k\,\rho^*_k(\xia)\,\V_k\nonumber\\
	&= - i\,\frac{\qa L}\ma\sumk k\left[\rho^*_k(\xia)\,\V_k - \rho_k(\xia)\,\V^*_k\right]\nonumber\\
	& = \frac{\qa L}\ma\sumk 2k\Imag\left[\rho^*_k(\xia)\,\V_k\right]\label{eom-fb}
\end{align}
The spatial charge density associated with a single particle is 
\begin{equation}
	\qa \sum_k \rho_k(\xia)\,\bv_k(x)
\end{equation}
and the corresponding total charge is
\begin{align}
	\qa \sum_k \rho_k(\xia)\intdxL \bv_k(x) &= 	\qa \sum_k \rho_k(\xia)\,L\,\delta_{k0}\nonumber\\
	&= \qa\,L\, \rho_0(\xia)\nonumber\\
	&= \qa\,L\intdxL S(x-\xia)\,\frac1L\nonumber\\
	&= \qa\,.
\end{align}
Thus, regardless of the number of modes retained, the charge associated with each particle remains $\qa$.

The energy of this system is
\begin{equation}
	W_L = \frac\ma2\sump \wa\,\xida^2 + \frac{L}{4\pi}\sumk k^2\V_k\,\V^*_k\,.
\end{equation}
Using the equations of motion we have
\begin{align}
	\frac{dW_L}{dt} &= \ma \sump \wa\,\xida\,\xidda + \frac{L}{4\pi}\,\sumk k^2\left(\dot\V_k\,\V^*_k + \V_k\,\dot\V^*_k\right) \nonumber\\
	 & = -i\,\qa\,L\sump\sumk \wa\,k\,\xida\left(\rho^*_k\,\V_k - \rho_k\,\V^*_k\right) - i\,\qa\,L\sumk\sump k\,\xida\left(\rho_k\V^*_k - \rho^*_k\,\V_k\right) \nonumber\\
	 & = 0\,,
\end{align}
where we have used \eqref{poisson-fb} and \eqref{drhodxi} to find $\dot\V_k$.  From \eqref{P_cont} we have
\begin{align}
	\frac{dP}{dt} & = \ma\sump \wa\,\xidda \nonumber\\
	& = -i\,\qa\,L\sump\sumk \wa\,k\left[\rho^*_k(\xia)\,\V_k - \rho_k(\xia)\,\V^*_k\right] \nonumber\\
	& = -i\,\frac L{4\pi}\sumk k^3\left(\V^*_k\,\V_k - \V_k\,\V^*_k\right) \nonumber\\
	& = 0\,.
\end{align}
Thus the model using a truncated Fourier basis conserves both energy and momentum.  This is as expected since the spatial discretization does not introduce time-dependence into the Lagrangian and the
basis was specifically constructed to maintain spatial translation invariance.

Consider the same system as in Section \ref{spatial-grid} with $S$ given by \eqref{top-hat-S} where $h$ is an independent parameter.  Now \eqref{rho_k-fb} becomes
\begin{equation}
	\rho_k(\xia) = \frac1L\,e^{-ik\xia}\sinc\left(\tfrac12kh\right),
\end{equation}
where $\sinc x = \sin(x)/x$.  If, as before, we take a quasi-neutral plasma with a uniform ion density, then the ions only contribute to Poisson's equation for $k=0$.  Further, we see that $k=0$ does
not contribute to $\xidda$, and thus we are free to take $\V_0=0$.  With this form of $\rho_k$, the potential becomes
\begin{equation}
	\V_k = \frac{4\pi\,\qa}{k^2L}\sinc\left(\tfrac12kh\right)\sump \wa\,e^{-ik\xia},\quad k>0
	\label{eom-pot-fb}
\end{equation}
and the particle equation of motion becomes
\begin{align}
	\xidda &= \frac{\qa}\ma\sumk (-i\,k)\sinc\left(\tfrac12kh\right)\left[e^{ik\xia}\,\V_k - \,e^{-ik\xia}\,\V^*_k\right]\nonumber\\[4pt]
	&= \frac{\qa}\ma\sumk\sinc\left(\tfrac12kh\right)\left[e^{ik\xia}E_k + \,e^{-ik\xia}\,E^*_k\right]\nonumber\\[4pt]
	&= \frac{\qa}\ma\sum_k\sinc\left(\tfrac12kh\right)E_k\,e^{ik\xia}
	\label{eom-th-fb}
\end{align}
where $E_k = -i\,k\,\V_k$.  For the complete basis, by the convolution theorem, \eqref{eom-th-fb} is identical to \eqref{xi-EOM-Cont}.  While, due to the truncation, the
convolution theorem does not apply, we may still interpret the force in \eqref{eom-th-fb} as sampling the electric field over the effective spatial extent of the particle.

In this subsection we have derived a particle algorithm that preserves the time and space translational invariance of the Lagrangian and thus conserves both energy and momentum exactly.  Since the use
of grid in the reduction violates the spatial translational invariance, we were lead to use a continuous basis.  In the course of the derivation, we found that only one such basis exists, a (possibly
truncated) Fourier basis.  One may argue that the use of a Fourier limits the applicability of this algorithm; for example, restricting to systems with periodic boundary conditions or being unsuitable
for large-scale parallel simulations.  This result establishes, however, that simultaneous conservation of energy and momentum is indeed possible.\label{fb-summary}

\section{Noncanonical Hamiltonian formulation}
\label{BracketDerivation}

It is well known \cite{Morrison:1982aa,Weinstein-Morrison81,morrison:1980:383} that the Vlasov--Maxwell system possesses a Hamiltonian structure in terms of non-canonical field variables.
Specializing to the 1-D electrostatic case and treating the electric field $E$ as a dynamical variable, the Vlasov--Maxwell bracket \cite{morrison:1980:383,Morrison:1982aa} becomes
\begin{equation}
	\PB F,G; = \intdxp f \pb{\fd F,f;}{\fd G,f;} + 4\pi\qa \intdxp \frac{\partial f}{\partial p} \left(\fd F,E;\,\fd G,f; - \fd G,E;\,\fd F,f; \right),
	\label{EM_Continuous_Bracket}
\end{equation}
where $F$ and $G$ are any functionals of $f$ and $E$ and $\pb ab$ denotes the usual phase-space Poisson bracket:
\begin{equation}
	\pb ab = \frac{\partial a}{\partial x}\frac{\partial b}{\partial p} - \frac{\partial a}{\partial x}\frac{\partial b}{\partial p}\, .
\end{equation}
The Vlasov equation and the equations for the fields are obtained from this bracket and the Hamiltonian
\begin{equation}
	H = \frac1{2\ma}\intdx p^2\,f + \frac1{8\pi}\intdx E^2
	\label{H_Continuous_Bracket}
\end{equation}
as
\begin{align}
	\frac{\partial f}{\partial t} &= \PB f,H; = - \frac p{\ma}\, \frac{\partial f}{\partial x} - \qa\,\frac{\partial f}{\partial p}\,E\,, \label{Vlasov}\\[4pt]
	\frac{\partial E}{\partial t}&= \PB E,H; = - 4\pi \intdp \frac p{\ma}\, f = - 4\pi j.  \label{dot_E}
\end{align}
Poisson's equation is considered as an initial condition and is satisfied for all time as a consequence of \eqref{dot_E}. 

We use a reduction of the distribution function, which is identical to \eqref{Sum_f_i} but written in terms of momentum:
\begin{align}
	f(x, p, t) & = \sum_\alpha\fa(x, p, t) \nonumber \\
	&= \sum_\alpha\wa\,S[x - \xia(t)] \,\delta[p - \pia(t)].\label{Sum_f_i_p}
\end{align}
Consider a single $\fa$. The quantities $\wa$, $\xia$, and $\pia$, which denote the macro-particle weight, centroid, and momentum, may be expressed as:
\begin{align}
  \wa & = \intdxp \fa\,, \\[2pt]
  \xia & = \frac1\wa\intdxp x\,\fa\,, \\[2pt]
  \pia & = \frac1\wa\intdxp p\,\fa\,.
\end{align}
Therefore, they may be thought of as functionals of $\fa$.  To an arbitrary functional $F[f]$ there exists a corresponding function $\Ft(\wa, \xia, \pia)$ such that $\Ft(\wa, \xia, \pia) = F[f]$.
(Both $F$ and $\Ft$ can also be functionals of $E$; for the moment, we are only interested in their dependence on $f$).  Then a functional derivative of $F[f]$ may be found using the chain rule as
\begin{equation}
	\fd F,\fa; = \fd\wa,\fa;\,\frac{\partial \Ft}{\wa} + \fd\xia,\fa;\,\frac{\partial \Ft}{\xia} + \fd\pia,\fa;\,\frac{\partial \Ft}{\pia}\,.
\end{equation}

Evaluating the functional derivatives of $\wa$, $\xia$, and $\pia$
\begin{align}
  \fd\wa,\fa; & = 1\,,\nonumber \\[2pt]
  \fd\xia,\fa; & = \frac{x - \xia}\wa\,, \\[2pt]
  \fd\pia,\fa; & = \frac{p - \pia}\wa\,\nonumber\,
\end{align}
we find
\begin{equation}
	\fd F,\fa; = \frac{\partial \Ft}{\partial\wa} + \frac{x - \xia}\wa\,\frac{\partial \Ft}{\partial\xia} + \frac{p - \pia}\wa\,\frac{\partial \Ft}{\partial\pia}\,.
\end{equation}
Consider
\begin{align}
	\pb{\fd F,\fa;}{\fd G,\fa;} & =\pb
	{\frac{\partial \Ft}{\partial\wa} + \frac{x - \xia}\wa\,\frac{\partial \Ft}{\partial\xia} + \frac{p - \pia}\wa\,\frac{\partial \Ft}{\partial\pia}}
	{\frac{\partial \Gt}{\partial\wa} + \frac{x - \xia}\wa\,\frac{\partial \Gt}{\partial\xia} + \frac{p - \pia}\wa\,\frac{\partial \Gt}{\partial\pia}}\nonumber  \\[4pt]
	& = \frac{\partial \Ft}{\partial\xia}\frac{\partial \Gt}{\partial\pia}\pb{\frac{x - \xia}\wa}{\frac{p - \pia}\wa} 
	  + \frac{\partial \Ft}{\partial\pia}\frac{\partial \Gt}{\partial\xia}\pb{\frac{p - \pia}\wa}{\frac{x - \xia}\wa}\nonumber  \\[4pt]
	& = \frac{\partial \Ft}{\partial\xia}\frac{\partial \Gt}{\partial\pia}\pb{\frac{x}\wa}{\frac{p}\wa} 
	  + \frac{\partial \Ft}{\partial\pia}\frac{\partial \Gt}{\partial\xia}\pb{\frac{p}\wa}{\frac{x}\wa}\nonumber  \\[4pt]
	 & = \frac1{\wa^2} \left(\frac{\partial \Ft}{\partial\xia}\frac{\partial \Gt}{\partial\pia} - \frac{\partial \Ft}{\partial\pia}\frac{\partial \Gt}{\partial\xia}\right)\nonumber  \\[4pt]
	 & = \frac1{\wa^2} \pb \Ft\Gt_{\xi\pi}
\end{align}
where
\begin{equation}
	\pb ab_{\xi\pi} = \frac{\partial a}{\partial\xia}\frac{\partial b}{\partial\pia} - \frac{\partial a}{\partial\xia}\frac{\partial b}{\partial\pia}\,.
\end{equation}
The first terms in \eqref{EM_Continuous_Bracket} then become
\begin{align}
	\intdxp\fa \pb{\fd F,\fa;}{\fd G,\fa;} & = \intdxp\fa \frac1{\wa^2} \pb \Ft\Gt_{\xi\pi}\nonumber\\[4pt]
	 & =\frac1{\wa^2}\pb \Ft\Gt_{\xi\pi}\intdxp\fa\nonumber  \\[4pt]
	 & =\frac1{\wa}\pb \Ft\Gt_{\xi\pi}.\label{PB-cont-E-t1}
\end{align}
Now consider
\begin{align}
	\intdxp\frac{\partial\fa}{\partial p}\fd F,E;\,\fd G,\fa; & = - \intdxp\fa\fd F,E;\,\frac{\partial}{\partial p}\fd G,\fa;\nonumber\\[4pt]
	& = \intdxp\fa\fd F,E;\frac{\partial}{\partial p}\left(\frac{\partial \Gt}{\partial\wa} + \frac{x - \xia}\wa\,\frac{\partial \Gt}{\partial\xia} + 
	\frac{p - \pia}\wa\,\frac{\partial \Gt}{\partial\pia}\right) \nonumber  \\[4pt]
	& = \intdxp\fa\fd F,E;\,\frac1\wa\,\frac{\partial \Gt}{\partial\pia}\nonumber  \\[4pt]
	& = \frac{\partial \Gt}{\partial\pia}\,\frac1\wa\intdx\fd F,E;\intdp\fa\nonumber  \\[4pt]
	& = \frac{\partial \Gt}{\partial\pia}\intdx S(x-\xia)\,\fd F,E;\,,\label{PB-cont-E-t2}
\end{align}
where the first line follows from integration-by-parts and the fact that $\delta F/\delta E$ does not have $p$ dependence since $E$ is a function of $x$ only.  Combining \eqref{PB-cont-E-t1} and
\eqref{PB-cont-E-t2} with \eqref{EM_Continuous_Bracket} leads to the bracket:
\begin{equation}
	\PB F,G; = \frac1\wa \pb \Ft\Gt_{\xi\pi} + 4\pi\qa\intdx S(x - \xia)\left(\fd G,E;\frac{\partial \Ft}{\partial\pia} - \fd F,E;\frac{\partial \Gt}{\partial\pia}\right)\, .
	\label{PB-cont-E}
\end{equation}
We now extend this result to a collection of $\fa$.  We treat each $\fa$ as a separate species, which mandates that the only interaction between the various $\fa$ is through the mean field.  The
bracket is thus just the sum of the \eqref{PB-cont-E} over $\alpha$:
\begin{equation}
	\PB F,G; = \sump\frac1\wa \pb \Ft\Gt_{\xi\pi} + 4\pi\qa\sump\intdx S(x - \xia)\left(\fd G,E;\frac{\partial \Ft}{\partial\pia} - \fd F,E;\frac{\partial \Gt}{\partial\pia}\right)\, .
	\label{PB-s-cont-E}
\end{equation}

Under our reduction, the Hamiltonian becomes (hereafter we drop the tilde notation as it should be clear from the above calculation where a functional derivative or a partial derivative is taken)
\begin{equation}
	H = \frac1{2\ma}\,\sump\wa\,\pia^2 + \frac1{8\pi}\intdx E^2
\end{equation}
and the equations of motion are
\begin{align}
	\xida = \PB\xia,H; &= \frac\pia\ma \label{xi-dot-ham}\\
	\pida = \PB\pia,H; &= \qa \intdx S(x-\xia)E(x) \label{pi-dot-ham}\\
	\frac{\partial E}{\partial t}  = \PB\pia,E; &= - 4\pi\qa\sump\wa\,\frac\pia\ma\,S(x-\xia)\nonumber\\
	&= -4\pi\qa\sump\wa\,\xida\,S(x-\xia) = -4\pi\,j\,. \label{E-dot-ham}
\end{align}
Equations \eqref{xi-dot-ham} and \eqref{pi-dot-ham} are easily seen to be equivalent to \eqref{xi-EOM-Cont}.  Comparing the spatial derivative of \eqref{E-dot-ham} to the time derivative of
\eqref{Poisson-Cont}, we see that \eqref{E-dot-ham} and \eqref{Poisson-Cont} are indeed equivalent.  As in the Lagrangian case, the reduction from a continuous phase space distribution function does
not break energy or momentum conservation.  In the Hamiltonian setting, energy conservation follows from the antisymmetry of the Poisson bracket under $F\leftrightarrow G$ and hence is intrinsic to
the theory.

To complete the reduction to a finite degree-of-freedom model, we represent $E$ using a finite, discrete basis, $\Psi_k$ having $N_b$ elements as
\begin{equation}
	E(x,t) = \sumbi E_i(t)\Psi_i(x)\,,
	\label{EB_Basis}
\end{equation}
where
\begin{equation}
	E_i(t) = \sumbj\intdx E(x,t)M^{-1}_{ij}\Psi_j(x)
	\label{Ek}
\end{equation}
and
\begin{equation}
M_{ij} = \intdx\, \Psi_i(x)\,\Psi_j(x)\,.
\label{Mass_Matrix}
\end{equation}
When $\Psi_i(x)$ are a finite element basis, $M_{ij}$ is called the mass matrix.  From \eqref{Ek}, we have
\begin{equation}
	\fd E_i,E; = \sumbj M^{-1}_{ij}\Psi_j(x).
\end{equation}
Now, the $E_i$, through \eqref{EB_Basis}, provide a complete characterization of $E$ and thus any functional of $E$ can be written as a function of the $E_i$.  Consequently
\begin{align}
	\fd ,E; &= \sumbi\fd E_i,E;\,\frac\partial{\partial E_i}\nonumber\\
	&= \sumbij M^{-1}_{ij}\Psi_j(x)\,\frac\partial{\partial E_i}\,.
\end{align}
Using this expression, the bracket becomes
\begin{equation}
	\PB F,G; =\sump\frac1\wa \pb FG_{\xi\pi}+ 4\pi\qa\sumbij\sump\left(\frac{\partial G}{\partial E_i}\,\frac{\partial F}{\partial\pia} - \frac{\partial F}{\partial E_i}\frac{\partial G}{\partial\pia}\right)
	M^{-1}_{ij}\rho_j(\xia)\,,
\label{Full_Bracket_Discr}
\end{equation}
where $\rho_j(\xia)$ is defined by \eqref{deposition}.  
The reduction of the bracket is exact in the sense that given the representation of $f$ and $E$, [\eqref{Sum_f_i_p} and \eqref{EB_Basis}, respectively] the reduced bracket and full bracket, restricted
to functionals of the appropriate form, give the same result.  Consequently, the reduced bracket inherits the Jacobi identity~\cite{Shadwick:2012aa} (and all other properties) from the full bracket.

Using \eqref{EB_Basis} and \eqref{Mass_Matrix}, we can write the Hamiltonian as
\begin{equation}\label{H_Discr_Bracket}
	H = \frac1{2\ma}\,\sump\wa\,\pia^2 + \frac1{8\pi}\sumbij M_{ij}\,E_i\,E_j.
\end{equation}
The equations of motion are then
\begin{align}
	\xida & = \frac\pia\ma \label{xi-dot-ham-d}\\
	\pida & = \qa \sumbi E_i\,\rho_i(\xia) \label{pi-dot-ham-d}\\
	\dot E_k & = -4\pi\qa\sump\sumbj\wa\,\frac\pia\ma\,M^{-1}_{kj}\rho_j(\xia) = -4\pi\qa\sump\sumbj\wa\,\xida\,M^{-1}_{kj}\rho_j(\xia) = -4\pi j_k\,. \label{E-dot-ham-d}
\end{align}

To make a connection with the model based on finite differences (Sec.~\ref{Lagrangian_PIC}), note that multiplication by the matrix $M$ is equivalent to performing an integration.  For finite elements
constructed from Lagrange polynomials one may reduce the mass matrix to a diagonal form (a procedure known as ``lumping'') while preserving the accuracy of the approximation \cite{jensen:1996}.  If we
use linear finite elements on a grid with spacing $h$, lumping the mass matrix gives
\begin{equation}
	M_{ij} \longrightarrow h\,\delta_{ij}.\label{Lumped_M}
\end{equation}

\section{Examples}
\label{Examples}

In this section we present two examples illustrating some properties of the energy conserving models derived in this paper.  We begin with a benchmarking example: the linear growth rate of the
instability caused by a small electron beam of density $n_b$ propagating in a neutralizing background plasma of density $n_0$ (beam-plasma instability).  For small beam-to-plasma density ratio,
$(n_b/n_0) \ll 1$ [more precisely, the parameter $({n_b}/{2n_0} )^{1/3}$ must be small in this linear theory], the linear growth rate of this instability is given by:
\begin{equation}
	\gamma_L = \frac{\sqrt{3}}{2}\left(\frac{n_b}{2n_0} \right)^{1/3}\omega_p.
	\label{growth_rate}
\end{equation}

\begin{figure}[htb]
	\centering
	\includegraphics{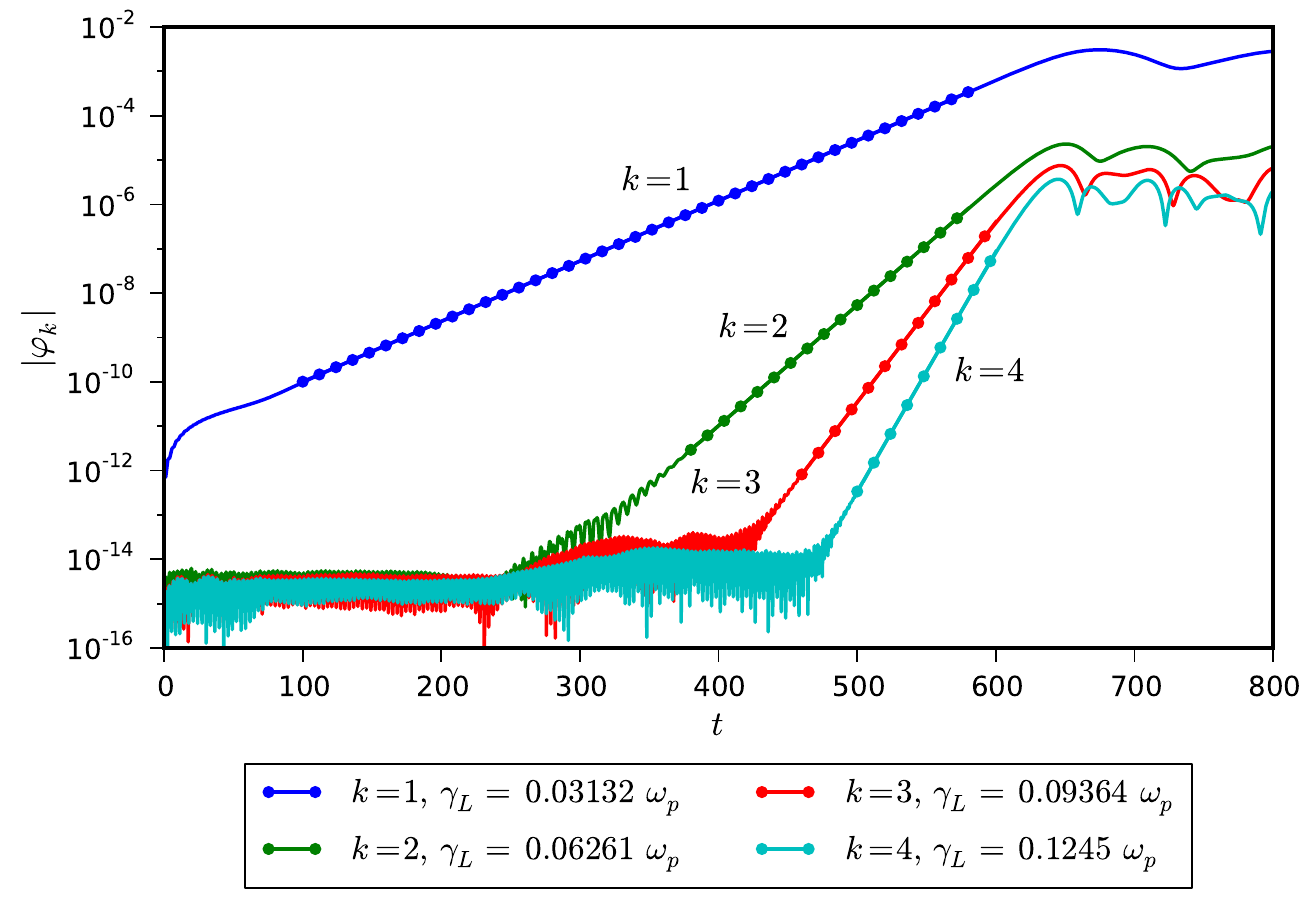}
	\caption{Linear growth and saturation of the first four harmonics in the beam--plasma problem computed using the truncated Fourier model, \eqref{eom-pot-fb} and \eqref{eom-th-fb}, for
	$n_b/n_0=10^{-4}$.  The analytical growth rate, \eqref{growth_rate}, gives $\gamma_L\approx0.03190$ for $k=1$.}
	\label{fig_fpic_linear_growth}
\end{figure}

All simulations are in dimensionless variables, where the time is measured in units of inverse plasma frequency, $\omega_p^{-1}$, momentum is measured in units of $m_e c$, potential is measured in
units of $m_e c^2/e$, and energy in units of $m_e c^3 n_0/\omega_p$ (assuming 1-D).  In the latter notation $m_e$ is the electron mass, $e$ is electron charge.  The system is assumed to be periodic
and its dimensionless size is $2\pi$.  In this way, the numerical growth rate is dimensionless while the physical growth rate is measured in units of $\omega_p$.  In Fig.~\ref{fig_fpic_linear_growth}
we show a simulation using the model of Sec.~\ref{TruncatedBasisDerivation}, \eqref{eom-pot-fb} and \eqref{eom-th-fb}.  The simulation was initialized by perturbing the beam density (position of beam
particles) at the wavelength of the first harmonic, and the velocity of the beam was matched to the plasma wave phase velocity, e.g., $v_{\rm beam} = 1$.  (To initialize the second harmonic, $k=2$,
the beam velocity would have to be set to $v_{\rm beam} = 1/2$, \textit{etc.}) The beam-to-plasma density ratio for this simulation is $10^{-4}$.  There are $300$ particles for each group of
particles, \textit{i.e.}, beam electrons, background (plasma) electrons, and plasma ions, as well as $128$ Fourier modes.  The plasma ions neutralize exactly both the beam and the plasma electrons,
which is achieved by an appropriate choice of particle weight (this assures the potential has zero bias).  The beam to plasma density ratio was also adjusted by an appropriate choice of beam and
background particles weight.

\begin{figure}[htbp]
	\centering
	\includegraphics[height=0.7\textheight]{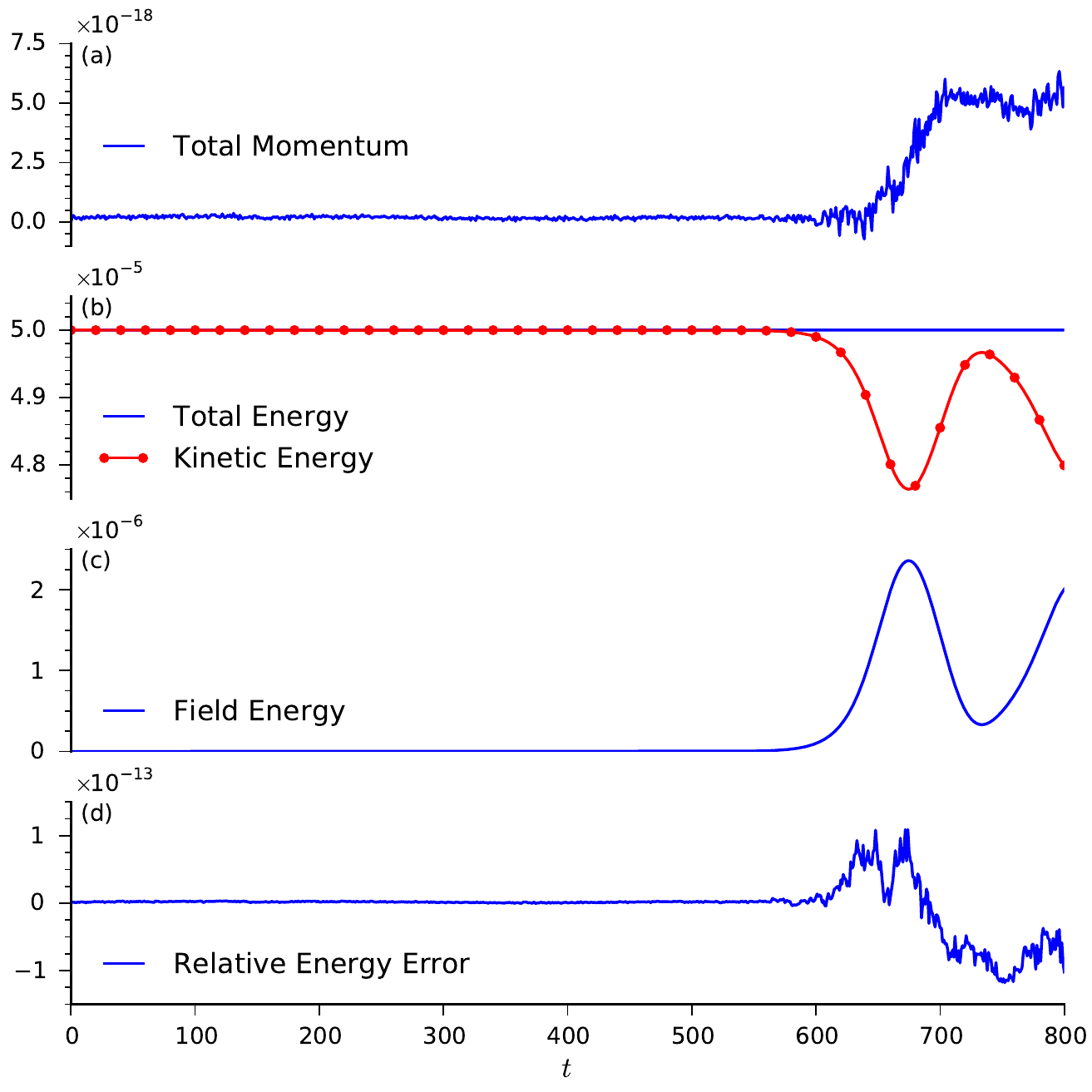}
	\caption{Momentum and energy balance for the simulation in Fig.~\ref{fig_fpic_linear_growth}.}
	\label{fig_fpic_energy_momentum}
\end{figure}

The numerical growth rate of the fundamental harmonic is approximately $0.03132$, which differs by less than $2\%$ from the theoretical value $\gamma_{L} = 0.03190$.  (The regions where the growth
rates are determined are indicated by dots.)  Better agreement can be achieved for smaller beam-to-plasma density ratios.  Also seen from this figure is that the next three harmonics grow sequentially
as a result of the non-linearity developing in the growth of the previous harmonics; \textit{i.e.,} the second harmonic is seeded by the non-linearity of the first harmonic when the quadratic term of the field
has grown sufficiently, etc.  In this scenario, linear growth rates of higher harmonics are multiples of the growth rate of the first harmonic independent of how well the numerical growth rate agrees
with formula (\ref{growth_rate}), as long as a clear linear stage exists; this is indeed the case in Fig.~\ref{fig_fpic_linear_growth}.  Energy and momentum balance are shown in
Fig.~\ref{fig_fpic_energy_momentum}.  Momentum is conserved to machine precision even in the time-discretized model, while energy conservation depends on the time integrator properties and time step
$\Delta t$.  To show the flexibility of the particle algorithm with respect to a choice of a time integration scheme, we chose a symplectic integrator of fourth order accuracy (the PEFRL algorithm of
Omelyan \textit{et al.}~\cite{omelyan_optimized_2002}).  For a choice of time step $\Delta t=0.01$, energy conservation is virtually perfect, at approximately $10^{-13}$ maximum relative error.

\begin{figure}[htb]
	\centering
	\includegraphics{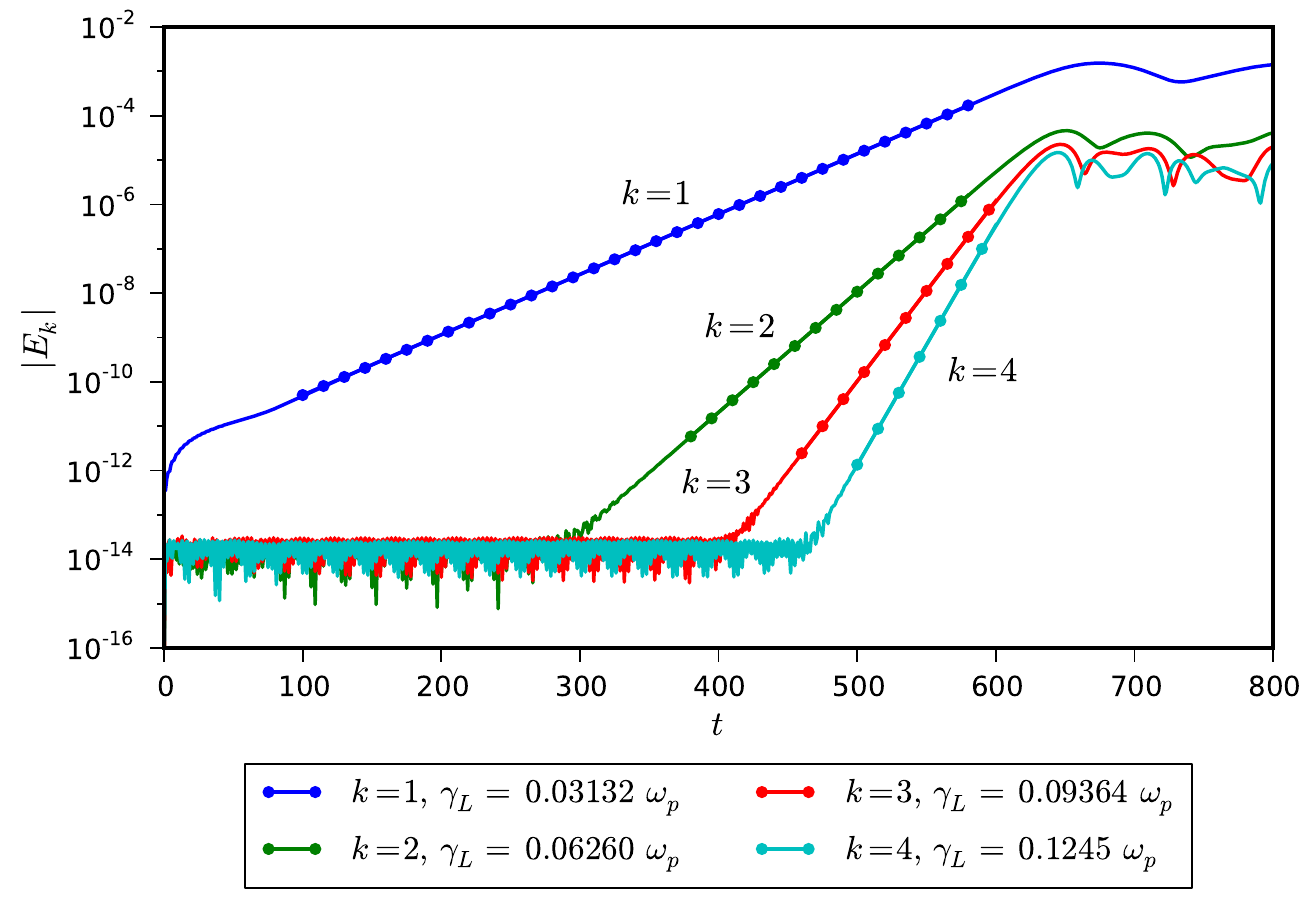}
	\caption{Linear growth and saturation of the first four harmonics in the beam--plasma problem computed using potential-based particle model, \eqref{xi-EOM} and \eqref{Poisson-d}, for
	$n_b/n_0=10^{-4}$ and the same simulation parameters as in Fig.~\ref{fig_fpic_linear_growth}.}
	\label{fig_pic_linear_growth}
\end{figure}

Similar results, shown in Fig.~\ref{fig_pic_linear_growth}, are obtained when the beam-plasma instability simulation is performed with the potential-based model, \eqref{xi-EOM} and \eqref{Poisson-d}.
(The regions where the growth rates are determined are indicated by dots and correspond to the same regions used in Fig.~\ref{fig_fpic_linear_growth}.)  The equations of motion were integrated with a
second-order Runge-Kutta method with time step $\Delta t=0.001$.  No time-splitting was used, \textit{i.e.}, all particle and field data are known at common points in time.  The number of grid points
was $2048$, the number of particles per cell was $4$, $\rho_{k}$ was cubic in $\xia$, corresponding to the shape $S_{1}$ (see Table~\ref{shapes}).  The growth rate of the first few harmonics is in
excellent agreement with the truncated Fourier series model.  Energy conservation for this model is also very good, with relative energy error of less that $0.6\%$ (not shown).

The examples of Figs.~\ref{fig_fpic_linear_growth}--\ref{fig_pic_linear_growth} demonstrate that energy conserving algorithms perform reliably in this benchmarking test and have low noise due to the
freedom to choose smooth particle shapes.

\begin{figure}[htb]
	\begin{center}
		\includegraphics{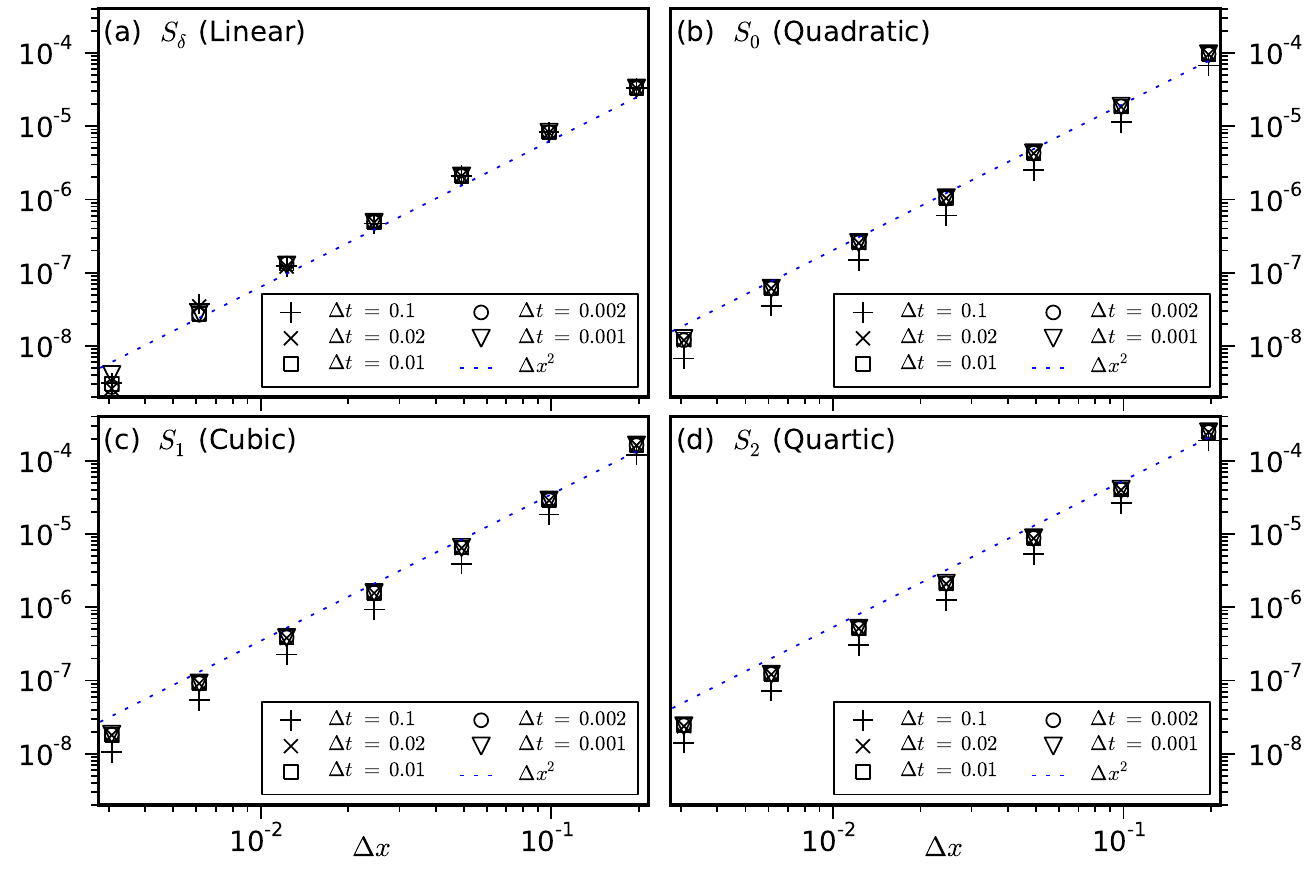}
	\end{center}
	\caption{Dependence of the $k=1$ growth rate in the beam--plasma problem on grid size for the potential-based particle model, \eqref{xi-EOM} and \eqref{Poisson-d}.  Plotted is the difference
	between the growth rate calculated using a given value of $\Delta x$ compared to the growth rate computed with very high spatial resolution for various values of $\Delta t$.  As the plot shows,
	the method is second-order, as expected by our approximation of $\LL$.  The panels are labeled with the particle shape $S$ and the resulting order of $\rho_{k}$ (see Table~\ref{shapes}).}
	\label{fig_lpic_field_energy_error_dx}
\end{figure}

Figure~\ref{fig_lpic_field_energy_error_dx} shows the dependence of the $k=1$ growth rate of the beam-plasma problem on the spatial resolution in the potential-based model \eqref{xi-EOM} and
\eqref{Poisson-d}.  Plotted is the difference between the growth rate calculated using a given value of $\Delta x$ compared to the growth rate computed with very high spatial resolution (and the
specified value of $\Delta t$.)  As expected from our approximation to $\LL$, growth rate is second order in the grid spacing, $\Delta x$, regardless of the particle shape or time-step.

\begin{figure}[htb]
	\begin{center}
		\includegraphics{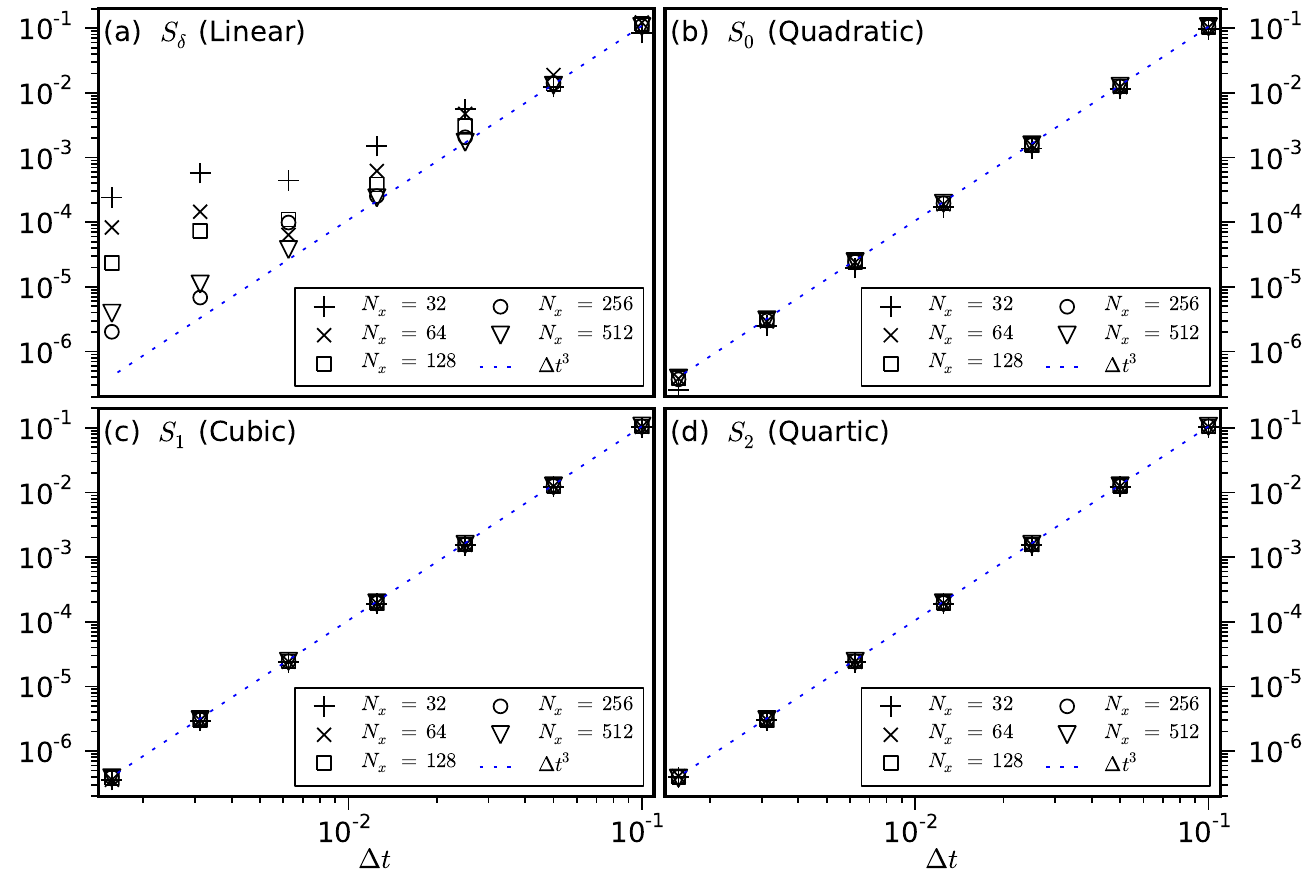}
	\end{center}
	\caption{Conservation of energy for the potential-based particle model, \eqref{xi-EOM} and \eqref{Poisson-d}.  The relative energy error is shown as a function of $\Delta t$ for various spatial
	resolutions and particle shapes.  As expected, the energy error depends only on the temporal discretization.  The equations of motion were solved with a second order Runge-Kutta integrator.  The
	panels are labeled with the particle shape $S$ and the resulting order of $\rho_{k}$ (see Table~\ref{shapes}).}
	\label{fig_lpic_potential_energy_error}
\end{figure}

\begin{figure}[htb]
	\begin{center}
		\includegraphics{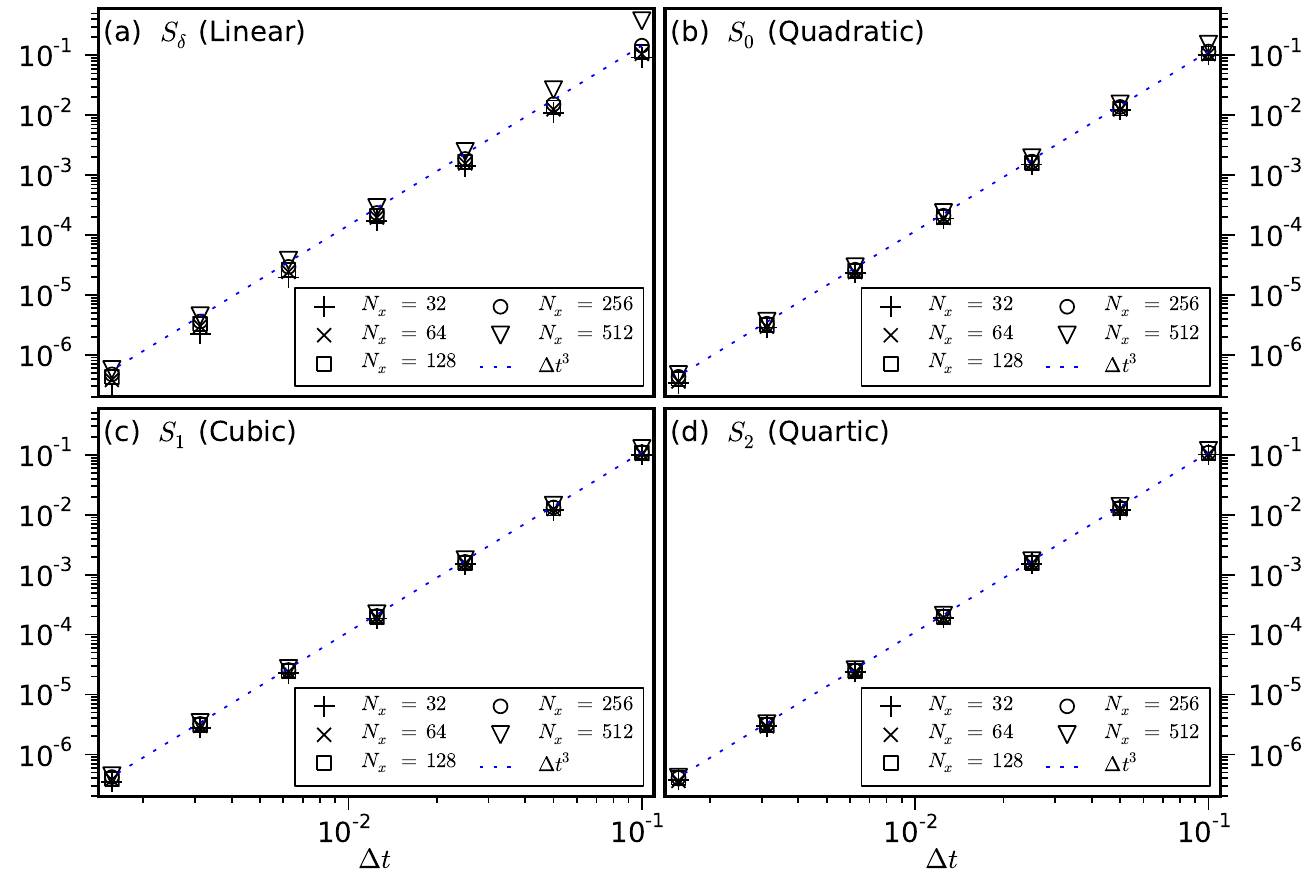}
	\end{center}
	\caption{Conservation of energy for the field-based particle model, \eqref{xi-dot-ham-d}--\eqref{E-dot-ham-d}).  The relative energy error is shown as a function of $\Delta t$ for various spatial
	resolutions and particle shapes.  As expected, the energy error depends only on the temporal discretization.  The equations of motion were solved with a second order Runge-Kutta integrator.  The
	panels are labeled with the particle shape $S$ and the resulting order of $\rho_{k}$ (see Table~\ref{shapes}).}
	\label{fig_lpic_field_energy_error}
\end{figure}

\begin{figure}[htb]
	\begin{center}
		\includegraphics{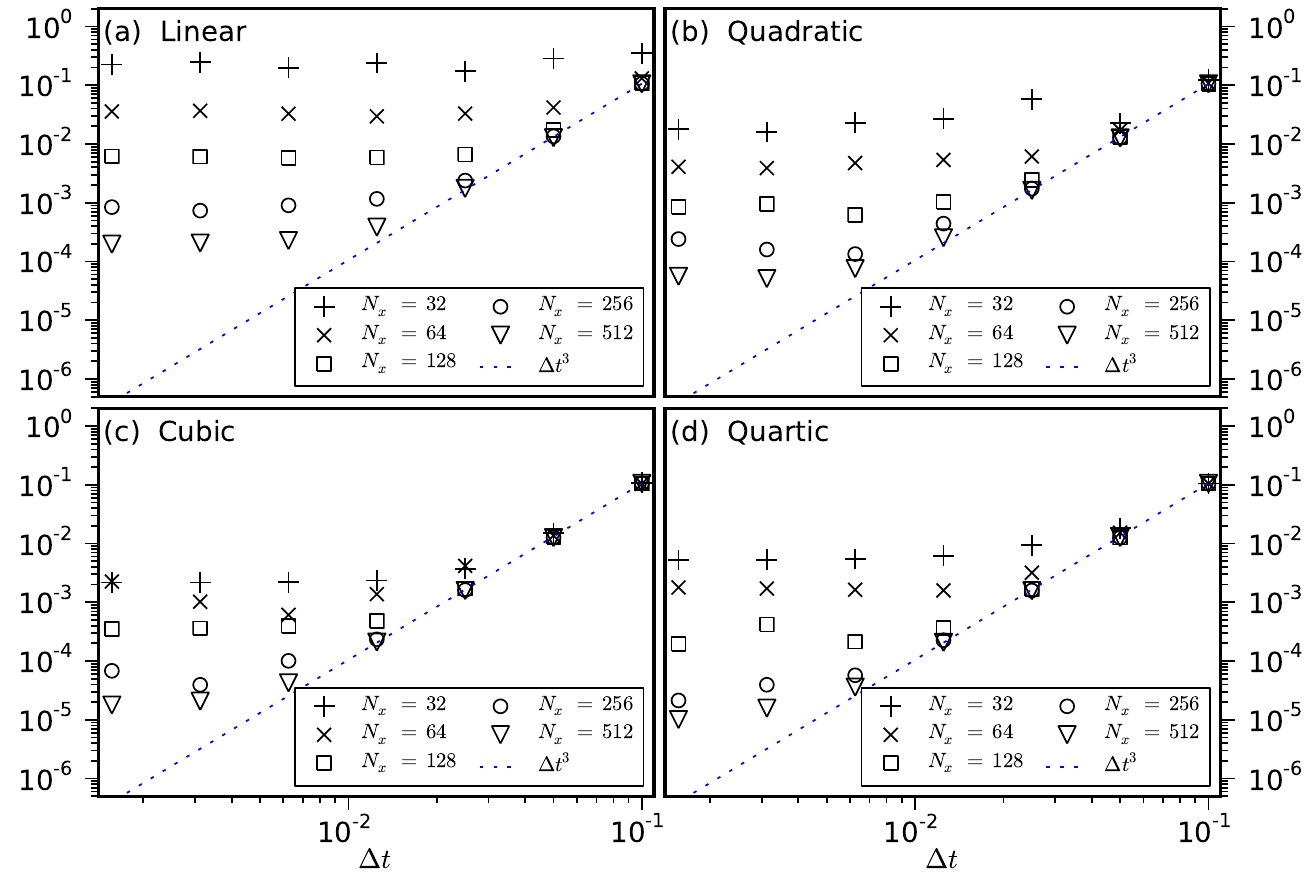}
	\end{center}
	\caption{Conservation of energy for a standard (momentum conserving) PIC algorithm.  The relative energy error is shown as a function of $\Delta t$ for various spatial resolutions and particle
	shapes.  The equations of motion were solved with a second order Runge-Kutta integrator.  The panels are labeled with the order of the charge-deposition/force-interpolation spline used; see
	Ref.\cite{Hockney88}.}
	\label{fig_pic_field_energy_error}
\end{figure}

\afterpage{\clearpage}

The next example illustrates an important property of these energy conserving algorithms, namely, that energy conservation depends solely on the properties of the time discretization.  We consider a
linear plasma oscillation, with electric field amplitude of $0.1$ and integrate the equations of motion with a second-order Runge-Kutta method.  The relative energy error at $t=400$ is plotted against
the time step $\Delta t$ for various particle shapes (see Table \ref{shapes}) and grid resolutions.  Figures~\ref{fig_lpic_potential_energy_error} and \ref{fig_lpic_field_energy_error} show the
relative error for the potential-based, \eqref{xi-EOM} and \eqref{Poisson-d}, and field-based, \eqref{xi-dot-ham-d}--\eqref{E-dot-ham-d}, energy conserving algorithms, respectively.  The scaling with
time step for all particle shapes is $\sim O(\Delta t^3)$; the exception is only for the potential-based method with linear $\rho_{k}$.  In this case the force on a particle is discontinuous, i.e., it
has jumps as a particle moves from one cell to another.  Therefore, for the potential-based formulation linear particles are not recommended.  Interestingly, in the case of linear particles in the
field-based formulation, this deficiency does not show as strongly and the scaling has the same trend as for smoother particle shapes, Fig.~\ref{fig_lpic_field_energy_error}.  Simulations with the
truncated Fourier basis particle model exhibit similar behavior (not shown).  Figure~\ref{fig_pic_field_energy_error} shows energy conservation in the standard (momentum conserving) PIC algorithm.  As
expected, the relative energy error in these algorithms depends on both the time step as well as the spatial grid resolution.  Note that for the same particle smoothness and time step, the energy
conserving algorithms have a (much) smaller relative energy error for $\Delta t \lesssim 10^{-2}$; for smaller $\Delta t$, energy conservation in the PIC algorithm is limited by the maximum number of
grid points being fixed at $512$.  These examples combined with the example in Fig.~\ref{fig_lpic_field_energy_error_dx} demonstrate that our method has overall accuracy of second order in both time
and space.

\section{Conclusions}
\label{Conclusions}

We have derived time-explicit, energy-conserving algorithms based on two approaches: Lagrangian in terms of potentials and a Hamiltonian with a non-canonical Poisson bracket in terms of fields.  These
models are derived without specifying any particular spatial or time discretization scheme, accuracy, or particle shape.  Our general method allows the Lagrangian-based derivation to relax a number of
restrictions imposed previously.  Continuous quantities are reduced by performing either a grid reduction (i.e., finite differences) or truncated bases.  When a grid reduction is used, mass matrices
do not appear, which decreases computational load and improves the efficiency of memory usage.  The important role of the particle shape and its relation to force interpolation is exhibited.  A
relaxed choice of particle shape helps decrease numerical noise in energy-conserving algorithms.  A Hamiltonian derivation is presented here for the first time.  The method uses a reduction of both
the Hamiltonian and the non-canonical Poisson bracket.  Since its formulation is in terms of fields, it avoids solving Poisson's equation.  A model conserving both energy and momentum is derived and
the conditions that make this possible are described.  Its derivation uses the relation between conservation laws and Lagrangian symmetries, thus emphasizing the power of variational principles.
Numerical benchmarking confirms the improvements in our algorithms.  It is shown that conservation of energy in all particle models derived here only depends on the accuracy of time integration; in
comparison, energy conservation in PIC depends on both the grid spacing and time integration accuracy.  We restricted our discussion to the case of a one-dimensional, nonrelativistic, unmagnetized,
electrostatic plasma.  The generalization to three dimensional, relativistic, electromagnetic plasmas is straightforward and will be presented elsewhere.  It is shown how to increase overall accuracy
(in space and time) beyond second order.

\section*{Acknowledgments}

This work was supported in part by the US DoE under contract number DE-FG02-08ER55000 and by the University of Nebraska Atomic, Molecular, Optical, and Plasma Physics Program of Excellence.

\appendix
\section{Particle Shapes}
\label{particle-shapes}

\begin{figure}[htb]
	\centering
	\includegraphics{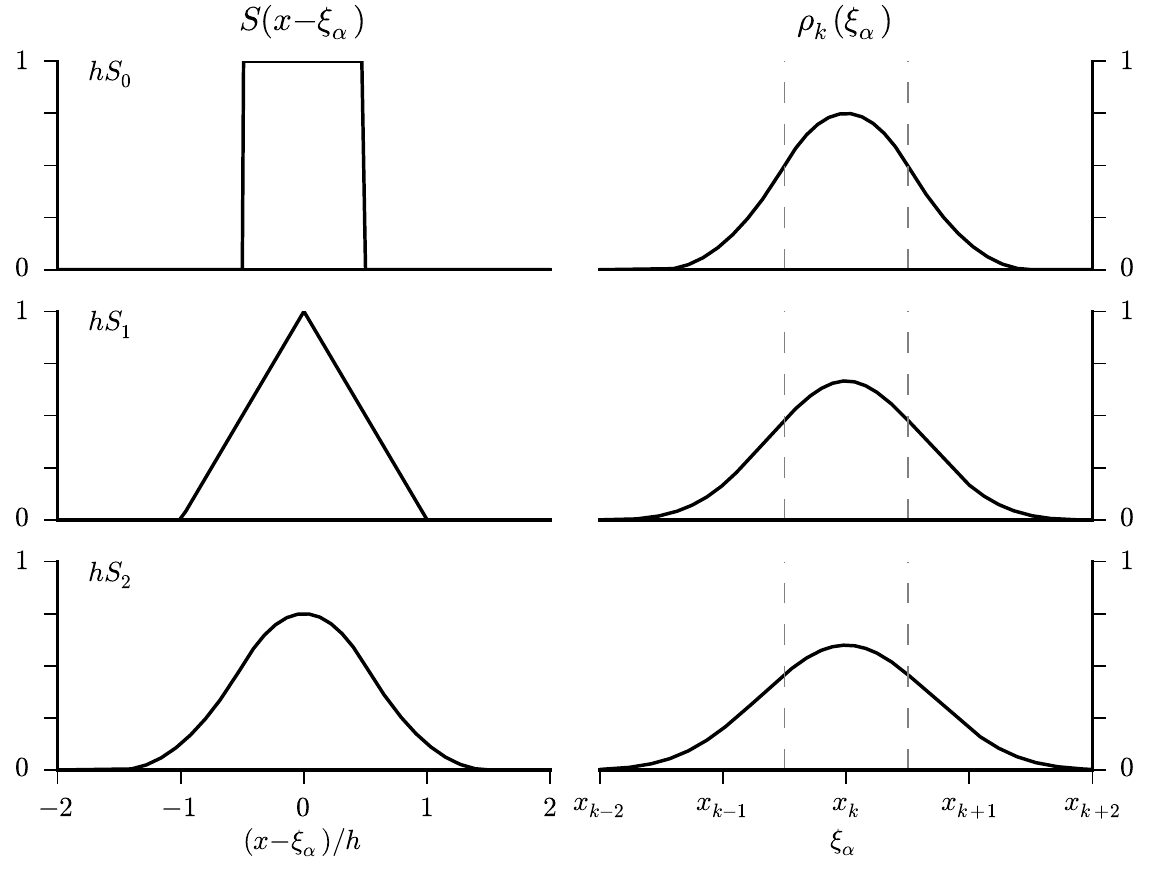}
	\caption{Particle shapes and corresponding charge deposition.  Charge deposition corresponding to various particle shapes $S(x - \xia)$.  For a given particle position $\xia$, $x_k$ is the nearest
	grid point.  We show the entire range of $\xia$ for which $\rho_k$ is non-zero so that the effective particle shape is apparent.  The range of $\xia$ for which $|x_k - \xia| < h/2$ is indicated by
	the dashed grey lines.  See Table \ref{shapes} for the definitions of $S_n$.}
	\label{shapes-fig}
\end{figure}

We present charge-deposition rules $\rho_i(\xia)$ for a variety of particle shapes based on \eqref{deposition} where the interpolation uses linear finite-elements; see Table \ref{shapes} and Figure
\ref{shapes-fig}.  The shapes $S_0$, $S_1$ and $S_2$ correspond to the usual PIC particle shapes scaled by $1/h$.  (The standard PIC definition normalizes the particle shape to have area $h$ whereas we
normalize our shapes to unity.)  This helps to explain the poor energy conservation observed with linear deposition (see Figures \ref{fig_lpic_potential_energy_error} and
\ref{fig_pic_field_energy_error}); such deposition correspond delta-function macro-particles.  As a result, any deposition scheme used should be at least quadratic in $\xia$ (\textit{i.e.}, at least
$C^2$).  All of these charge deposition rules are second-order accurate; higher-order accuracy is obtained by with a correspondingly higher order interpolation method, \textit{e.g.} using quadratic
finite-elements would lead to a third order accurate interpolation.

\begin{figure}[hbt]
	\centering
	\includegraphics{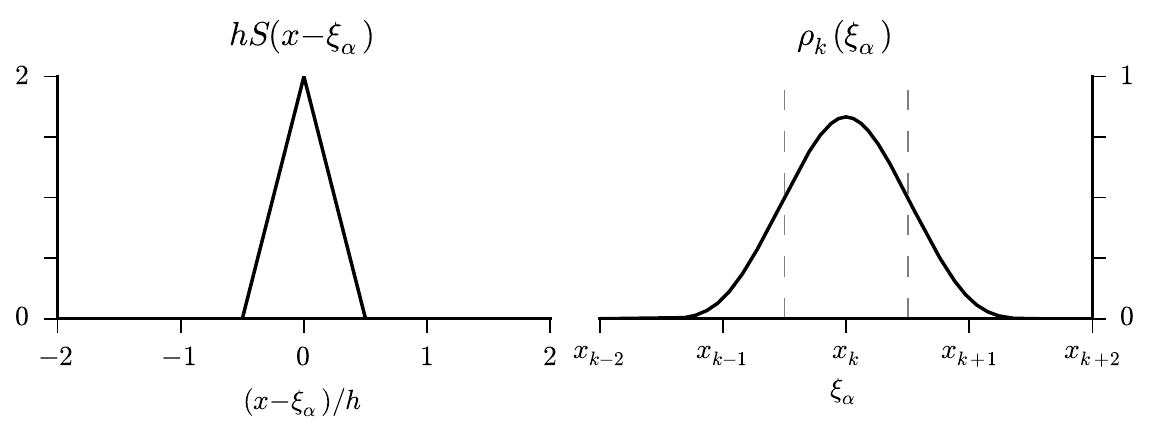}
	\caption{Charge deposition corresponding to various particle shapes $S(x - \xia)$.  For a given particle position $\xia$, $x_k$ is the nearest
	grid point.  We show the entire range of $\xia$ for which $\rho_k$ is non-zero so that the effective particle shape is apparent.  The range of $\xia$ for which $|x_k - \xia| < h/2$ is indicated by
	the dashed grey lines.  See \eqref{half-tent} for the definitions of $S_n$.}
	\label{half-tent}
\end{figure}

As discussed in Section~\ref{Lagrangian_PIC} in our formulation, there is no requirement for the particle size to be tied to the grid spacing.  To demonstrate this freedom, consider
\begin{equation}
	S(x-\xia) = \frac1h
	\begin{cases}
		\displaystyle1 - \frac2h\,|x-\xia|&|x-\xia|\le \tfrac12h,\\[8pt]
		\displaystyle0&\text{otherwise.}
	\end{cases}
	\label{S-ht}
\end{equation}
From Eq.~\eqref{deposition}, using linear finite elements for interpolation, and with the same definition of $\Delta$ as above [$\Delta = (\xia - x_k)/h$], we have for $\Delta > 0$
\begin{equation}
	\begin{aligned}
		\rho_{k-1}           &= \frac23\left(\frac12 - \Delta\right)^3,\\[4pt]
		\rho_{k\phantom{-1}} &= \frac56 - 2\,\Delta^2 + \frac43\,\Delta^3,\\[4pt]
		\rho_{k+1}           &= \frac1{12} + \frac12\,\Delta + \Delta^2 - \frac23\,\Delta^3,
	\end{aligned}
	\label{rho-ht-1}
\end{equation}
and for $\Delta < 0$ 
\begin{equation}
	\begin{aligned}
		\rho_{k-1}           &= \frac1{12} - \frac12\,\Delta + \Delta^2 + \frac23\,\Delta^3,\\[4pt]
		\rho_{k\phantom{-1}} &= \frac56 - 2\,\Delta^2 - \frac43\,\Delta^3,\\[4pt]
		\rho_{k+1}           &= \frac23\left(\frac12 + \Delta\right)^3.
	\end{aligned}
	\label{rho-ht-2}
\end{equation}
This particle shape results in a cubic deposition scheme (\textit{i.e.} the $\rho_k$ are $C^3$ in $\xia$), involving only three grid points (or 2 cells); see Figure~\ref{half-tent}.  In contrast, the
usual PIC cubic deposition involves four grid-points corresponding to a particle three cells in extent. Because of the linear finite elements used, this still produces second order accurate force interpolation.

\begin{table}[p]
	%
	%
	\def\heading#1{\hline\multicolumn{2}{@{}l}{\Spacer#1}\\[3pt]\hline}

	\def\DeltaP{\spacer\DeltaC>}
	\def\DeltaN{\DeltaC<}
	\def\DeltaC#1{\hbox{\footnotesize\underline{\strut$\Delta #1 0$}}}
	\def\pmo{\phantom{-1}}
	\def\spacer{\vrule width 0pt height9pt depth 0pt}
	\def\Spacer{\vrule width 0pt height13pt depth 0pt}
	\centering
	\begin{tabular}{@{}>{$\displaystyle}c<{$}|>{$\displaystyle}l<{$}}
		\multicolumn{1}{c}{$S(x-\xia)$}&\multicolumn{1}{c}{$\rho_i(\xi_\alpha)$}\\[2pt]
		\hline\hline
		S_\delta = \delta(x-\xia)&
		\begin{aligned}
			\DeltaP\\[2pt]
			\rho_{k-1}   & = 0\\
			\rho_{k\pmo} & = 1 - \Delta\\
			\rho_{k+1}   & = \Delta\\[6pt]
			\DeltaN\\[2pt]
			\rho_{k-1}   & = -\Delta\\
			\rho_{k\pmo} & = 1 - \Delta\\
			\rho_{k+1}   & = 0\\[3pt]
		\end{aligned}
		\\
		\hline
		S_0 = \frac1h
		\begin{cases}
			\displaystyle1&|x-\xia|\le\frac12h,\\[8pt]
			\displaystyle0&\text{otherwise.}
		\end{cases}
		&
		\begin{aligned}
			\rho_{k-1}   & = \tfrac12\left(\Delta - \tfrac12\right)^2\Spacer\\[3pt]
			\rho_{k\pmo} & = \tfrac34-\Delta^2\\[3pt]
			\rho_{k+1}   & = \tfrac12\left(\Delta + \tfrac12\right)^2\\[4pt]
		\end{aligned}
		\\
		\hline
		S_1 = \frac1h
		\begin{cases}
			\displaystyle1 - \frac{|x-\xia|}h&|x-\xia|\le h,\\[8pt]
			\displaystyle0&\text{otherwise.}
		\end{cases}
		&
		\begin{aligned}
			\DeltaP\\[4pt]
			\rho_{k-2}   & = 0\\
			\rho_{k-1}   & = \tfrac16 - \tfrac12\,\Delta + \tfrac12\,\Delta^2 - \tfrac16\,\Delta^3\\[3pt]
			\rho_{k\pmo} & = \tfrac23 - \Delta^2 + \tfrac12\,\Delta^3\\[3pt]
			\rho_{k+1}   & = \tfrac16 + \tfrac12\,\Delta + \tfrac12\,\Delta^2 - \tfrac12\,\Delta^3\\[3pt]
			\rho_{k+2}   & = \tfrac16\,\Delta^3\\[6pt]
			\DeltaN\\[4pt]
			\rho_{k-2}   & = -\tfrac16\,\Delta^3\\[3pt]
			\rho_{k-1}   & = \tfrac16 - \tfrac12\,\Delta + \tfrac12\,\Delta^2 + \tfrac12\,\Delta^3 \\[3pt]
			\rho_{k\pmo} & = \tfrac23 - \Delta^2 - \tfrac12\,\Delta^3\\[3pt]
			\rho_{k+1}   & = \tfrac16 + \tfrac12\,\Delta + \tfrac12\,\Delta^2 + \tfrac16\,\Delta^3\\[3pt]
			\rho_{k+2}   & = 0\\[4pt]
		\end{aligned}
		\\
		\hline
		\begin{aligned}
		&S_2 = \frac1h\times \vrule width 0pt height15pt\\[4pt]
		&\begin{cases}
			\displaystyle \tfrac34 - \left(\tfrac{x-\xia}h\right)^2           &            |x-\xia| \le \frac12h,\\[8pt]
			\displaystyle \tfrac12\left(\tfrac32 - \tfrac{|x-\xia|}h\right)^2 & \frac12h < |x-\xia| \le \frac32h,\\[8pt]
			\displaystyle 0                                                   & \text{otherwise.}
		\end{cases}
		\end{aligned}
		&
		\begin{aligned}
			\rho_{k-2}   & = \tfrac1{24}\left(\Delta - \tfrac12\right)^4\Spacer\\[2pt]
			\rho_{k-1}   & = \tfrac{19}{96} - \tfrac{11}{24}\,\Delta + \tfrac14\,\Delta^2 + \tfrac{1}{6}\,\Delta^3 - \tfrac{1}{6}\,\Delta^4\\[2pt]
			\rho_{k\pmo} & = \tfrac{115}{192} - \tfrac58\,\Delta^2 + \tfrac14\,\Delta^4\\[2pt]
			\rho_{k+1}   & = \tfrac{19}{96} + \tfrac{11}{24}\,\Delta + \tfrac14\,\Delta^2 - \tfrac{1}{6}\,\Delta^3 - \tfrac{1}{6}\Delta^4\\[2pt]
			\rho_{k+2}   & = \tfrac1{24}\left(\Delta + \tfrac12\right)^4\\[4pt]
		\end{aligned}
		\\
		\hline
	\end{tabular}
	\caption{Charge deposition corresponding to various particle shapes $S$.  For a given particle position $\xia$, let $k$ be the nearest grid point and $\Delta = (\xia - x_k)/h$.  All $\rho_i$ other
	than those listed are identically zero.}	
	\label{shapes}
\end{table}

\section{A fluid--kinetic hybrid model}
\label{HybridModel}

Under certain conditions, plasma electrons, plasma ions, or both, may be well approximated by (charged) fluids.  Typically in hybrid models, one species is described as a fluid while the other as
particles (i.e., kinetically).  Here we describe a single species with a fluid--kinetic hybrid and make no assumptions about the inter-mixing of the fluid and kinetic elements.  A prototypical system
is a low-charge electron beam propagating in a cold, quasi-neutral plasma.  Only the beam (\textit{i.e.} the tail of the distribution) needs to be treated in a fully kinetic manner.  The bulk plasma
can be represented as a fluid.  When the bulk plasma thermal velocity is small compared to both the velocity of the electron beam and the phase velocities of plasma waves arising from the beam-plasma
interaction, we may take the fluid to be cold.

It is computationally advantageous to use such splitting when the kinetic population of the plasma is small compared to its fluid-like population.  The numerical noise in the hybrid description can be
much lower compared to the kinetic description of the entire plasma.  In addition, there can be a large computational speedup due to using a fluid description for the larger fraction of the plasma since
only time advance of gridded quantities is required.

For concreteness, we assume stationary ions and mobile electrons.  The cold electron fluid distribution function is approximated as a delta function in velocity space
\begin{equation}
	f({\bf x},{\bf v},t) = n({\bf x},t)\,\delta[{\bf v} - \Dt{\bf x}(t)]\,,\label{ColdFluid_f}
\end{equation}
where $n({\bf x},t)$ is the fluid density, $D_t{\bf x}(t)$ is the velocity of a Lagrangian fluid element ($\Dt = \partial_t + u_i\,\partial/\partial\tilde{x}_i$ being the convectional derivative), and
$\mathbf u$ is the Eulerian fluid velocity.  We express the fluid velocity in terms of the velocity potential $\vpot$ and the Clebsch variables $\alpha$ and $\beta$ as ${\bf u} = \nabla\vpot +
\alpha\nabla\beta$ with $\nabla\times\mathbf{u} = \nabla\alpha\times\nabla\beta$.  The kinetic electron distribution is given by expression \eqref{Sum_f_i} and the complete distribution function is
\begin{equation}
	f({\bf x},{\bf v},t) = n({\bf x},t)\delta[{\bf v} - \Dt{\bf x}(t) ] + \sump \wa S[{\bf x}-\xia(t)] \,\delta[{\bf v}-\xida(t)].\label{Hybrid_f}
\end{equation}
The reduction proceeds as before, by substitution of \eqref{Hybrid_f} into Low's Lagrangian, \eqref{Lows_Lagrangian}.  Since the Lagrangian is linear with respect to the distribution function we can
consider separately the fluid and kinetic contributions to $f$ simply adding the resulting Lagrangians.  The reduction of the kinetic contribution is identical to Section \ref{Lagrangian_PIC} and the
reduced Lagrangian (without the field contribution) is given by
\begin{equation}
	\Lp = \frac\ma2\sump \wa\,\xid^2_\alpha - \qa \sump\sumgi \wa \rho_i(\xia)\,\V_i\,.
\end{equation}
Below we consider only the fluid and field contributions
\eqref{L_Cont} together. At the end we add the particle, fluid, and field contributions.

Again for simplicity, we specialize to a non-relativistic, electrostatic system.  One may formulate the fluid variational principle in Eulerian picture using the fluid density, velocity potential, and
the Clebsch variables as independent variables \cite{de_veubeke}:
\begin{equation}
	\Lf = - \ma\intD{\meas{^3 x}}n \left[\frac12\left(\nabla\vpot + \alpha\nabla\beta\right)^2 + \dot{\vpot} + \alpha\,\dot\beta\right] + \frac{1}{8\pi}\intD{\meas{^3 x}}\left(\nabla
	\V\right)^2 - \qa\intD{\meas{^3 x}}n\,\V .
	\label{L_fl1}
\end{equation}
The second (field) term in \eqref{L_fl1} must be included only once in the final Lagrangian of particles and fluid.  As a further simplification, we develop the fluid model in one spatial dimension;
we also assume that ions form a uniform (immobile) background with density $n_0$.  In one spatial dimension, $\alpha=\beta=0$ and the fluid experiences electrostatic force through the electrostatic
potential $\V$.  The Lagrangian for this case is
\begin{equation}
	\Lf = -\ma\intdx n \left[\frac12\left(\nabla\vpot\right)^2 + \dot\vpot\right] + \frac1{8\pi} \intdx\left(\nabla\V\right)^2 - \qa\intdx\> n\,\V - \qi\,\nion\intdx\V\,.\label{L_fl}
\end{equation}
To proceed with the reduction, we first choose a linear finite element basis and expand all dependent variables. We will see, this approach leads to mass matrices, which need inversion. Then we use grid-based reduction, which eliminates mass matrices altogether, and therefore such numerical model has computational advantage. (Other choices, such as a truncated Fourier basis, may be more appropriate for some applications; nonetheless, the derivation proceeds along similar lines.)  

The fluid variables are represented as
\begin{equation}\label{FE_Fluid_Vars}
	\begin{aligned}
		\vpot(x,t) &= \sumgi \vpot_i(t)\,\feL_i(x)\,,\\[2pt]
		n(x,t)        &= \sumgi n_i(t)\,\feL_i(x)\,,\\[2pt]
		\V(x,t)       &= \sumgi \V_i(t)\,\feL_i(x)\,.
	\end{aligned}
\end{equation}
Substitution of \eqref{FE_Fluid_Vars} into the fluid Lagrangian \eqref{L_fl} gives
\begin{multline}
	\Lf = - \frac12\,\ma\!\!\!\sumg{i,j,k}F_{ijk}\vpot_i\vpot_j\,n_k - \ma\sumgij M_{ij}\dot{\vpot}_i\,n_j\\
	- \frac h{8\pi}\sumgij K_{ij} \V_i \V_j - \qa\sumgij M_{ij}\,\V_i\,n_j - \qi\,h\,\nion\sumgi\V_i,
	\label{FE_L_fl}
\end{multline}
where $h$ is the grid spacing,
\begin{equation}
	F_{ijk} = \intdx\frac{d\feL_i(x)}{dx}\,\frac{d\feL_j(x)}{dx}\,\feL_k(x)\,,
	\label{Fijk}
\end{equation}
and $K_{ij}$, and $M_{ij}$ are given by \eqref{Kij} and \eqref{Mass_Matrix}, respectively.  (We note that the coefficients $F_{ijk}$ are symmetric in $i$ and $j$ but {\it not} symmetric in all three
indices.)

This is a finite degree-of-freedom Lagrangian and thus requiring the action to be stationary, leads to the usual Euler--Lagrange equations
\begin{equation}
	\begin{aligned}
		\frac{d}{dt}\frac{\partial \Lf}{\partial\dot\vpot_k} - \frac{\partial\Lf}{\partial\vpot_k} & = 0\,,  \\[3pt]
		\frac{d}{dt}\frac{\partial \Lf}{\partial\dot\V_k} - \frac{\partial\Lf}{\partial\V_k} & = 0\,,  \\[3pt]
		\frac{d}{dt}\frac{\partial \Lf}{\partial\dot n_k} - \frac{\partial\Lf}{\partial n_k} & = 0\,.
	\end{aligned}
\end{equation}
The Euler--Lagrange equation for $n_l$ yields the cold fluid momentum equation:
\begin{equation}
	-\frac12\,\ma\sumgij F_{ijl}\vpot_i\vpot_j - \sumgi M_{il}\dot{\vpot}_i - \qa\sumgi M_{il}\,\V_i = 0
\end{equation}
or
\begin{equation}
	\dot{\vpot}_l = -\frac\qa\ma\,\V_l -\frac12\sumg{i,j,k}F_{ijk}M^{-1}_{kl}\vpot_i\vpot_j.\label{FE_Momentum_Eq}
\end{equation}
The Euler--Lagrange equation for $\vpot_l$ gives the fluid continuity equation:
\begin{equation}
	-\ma\sumgi M_{lj}\dot{n}_j + \ma\sumg{j,k} F_{ljk}\vpot_j n_k = 0
\end{equation}
or
\begin{equation}
	\dot{n}_i = \sumg{l,j,k}M^{-1}_{il}F_{ljk}\vpot_j\,n_k.
	\label{FE_Continuity_Eq}
\end{equation}
Finally, the Euler-Lagrange equation for $\V_k$ gives Poisson's equation (without the particle contribution):
\begin{equation}\label{Poisson_fluid}
	\sumg j K_{ij}\V_j = -4\pi\,\frac\qa h\sumg j M_{ij}n_j - 4\pi\,\qi\,\nion.
\end{equation}

Eqs.~\eqref{FE_Momentum_Eq}--\eqref{Poisson_fluid} are a complete set of equation for a fluid description of a plasma.  To obtain the hybrid particle-fluid model, these fluid equations must be
supplemented by the particle equation of motion and the particle contribution to Poisson's equation:
\begin{eqnarray}
{\xidda} &=& - \frac{q_s}{m_s}\sumgi\frac{\partial\rho_i(\xia)}{\partial\xia} \V_i \, , \label{hybrid_ddot_xi} \\
\sumg j K_{ij}\V_j &=& - 4\pi\,\frac\qa h \sump\wa\rho_i(\xia)  - 4\pi\,\frac\qa h\sumg j M_{ij}n_j - 4\pi\,\qi\,\nion\,,  \label{hybrid_Poisson}
\end{eqnarray}
where $\rho_i(\xia)$ is given by \eqref{deposition}.  The complete set of hybrid fluid-particle equations is given by Eqs.~\eqref{FE_Momentum_Eq}, \eqref{FE_Continuity_Eq}, \eqref{hybrid_ddot_xi}, and
\eqref{hybrid_Poisson}.  The conserved energy is
\begin{equation}\label{hybrid_Energy}
W_{L} = \frac12\,\ma\sump\wa\xida^2 - \frac{1}{2}\,\ma\sumg{i,j,k} F_{ijk}\vpot_i\,\vpot_j\,n_k - \frac h{8\pi}\sumgij K_{ij}\V_i\V_j.
\end{equation}

We see the appearance of the inverse of the mass matrix in Eqs.~(\ref{FE_Momentum_Eq}) and (\ref{FE_Continuity_Eq}). Having to keep the (dense) inverse of the mass matrix may consume too much computer memory and make computation times longer. Therefore, we show that with the use of the grid-based reduction mass matrices to not appear and no matrix inversion is necessary. The grid-based reduction
simply tells us to use a numerical integration rule and finite differences to reduce the integrals of continuous quantities. Thus, all bi-linear combinations of continuous quantities reduce to sums over the grid index. With these rules, the Lagrangian (\ref{FE_L_fl}) becomes
\begin{multline}
	\widetilde{\Lf} = - \frac12\,\ma\!\!\!\sumg{i,j,k}\widetilde{F}_{ijk}\vpot_i\vpot_j\,n_k - \ma h\sumgi \dot{\vpot}_i\,n_i\\
	- \frac h{8\pi}\sumgij \widetilde{K}_{ij} \V_i \V_j - \qa h\sumgi \V_i\,n_i - \qi\,h\,\nion\sumgi\V_i,
	\label{FE_L_fl_grid-based}
\end{multline}
with
\begin{equation}
	\widetilde{F}_{ijk} = h D_{ij}D_{ik}.
	\label{Fijk_modified}
\end{equation}
$\widetilde{K}_{ij}$ was defined in (\ref{K_tilde}) and the finite differencing operator $D_{ij}$ may be chosen as the second order accurate centered differencing
$D_{ij}=(\delta_{i,i+1}-\delta_{i,i-1})/2h$.  It is now clear how to obtain the corresponding modified equations of motion: replace $M_{ij}$ by $h$ (and $M_{ij}^{-1}$ by $h^{-1}$) and use the
coefficients (\ref{Fijk_modified}) in place of (\ref{Fijk}).  The energy expression (\ref{hybrid_Energy}) must also be modified accordingly.  Total momentum for this hybrid model is not conserved due
to the use of grid.


\end{document}